\documentclass[12pt,a4paper,twoside]{report}
\usepackage[applemac]{inputenc}
\usepackage{amssymb}
\usepackage{amsmath}
\usepackage{amsopn}
\include{introduction.tex}
\newcommand{\LL}{\mathbb{L}}
\newcommand{\CC}{\mathbb{C}}

\newcommand{\AAA}{\mathbb{A}}

\newcommand{\KK}{\mathbb{K}}

\newcommand{\NN}{\mathbb{N}}

\newcommand{\PP}{\mathbb{P}}

\newcommand{\RR}{\mathbb{R}}

\newcommand{\frc}{\mathfrak{C}}

\newcommand{\frb}{\mathfrak{B}}

\newcommand{\frp}{\mathfrak{p}}
\newcommand{\frpp}{\mathfrak{P}}

\newcommand{\kaa}{\mathcal{A}}

\newcommand{\ktm}{\mathcal{T}(M)}
\newcommand{\kcc}{\mathcal{C}}

\newcommand{\kD}{\mathcal{D}}
\newcommand{\kE}{\mathcal{E}}
\newcommand{\kf}{\mathcal{F}}

\newcommand{\kh}{\mathcal{H}}
\newcommand{\kI}{\mathcal{I}}
\newcommand{\kj}{\mathcal{J}}
\newcommand{\kK}{\mathcal{K}}
\newcommand{\kL}{\mathcal{L}}
\newcommand{\mm}{\mathcal{M}}

\newcommand{\kO}{\mathcal{O}}
\newcommand{\kP}{\mathcal{P}}
\newcommand{\kQ}{\mathcal{Q}}
\newcommand{\kR}{\mathcal{R}}
\newcommand{\kS}{\mathcal{S}}

\newcommand{\kT}{\mathcal{T}}

\newcommand{\gb}{\beta}
\newcommand{\gd}{\delta}
\newcommand{\eps}{\varepsilon}

\newcommand{\gG}{\Gamma}

\newcommand{\gl}{\lambda}

\newcommand{\gO}{\Omega}
\newcommand{\gf}{\varphi}

\newcommand{\gF}{\Phi}
\newcommand{\gr}{\varrho}
\newcommand{\gs}{\sigma}

\newcommand{\gt}{\tau}
\newcommand{\gtt}{\theta}

\newcommand{\tm}{\subseteq}

\newcommand{\lm}{\emptyset}
\newcommand{\∞}{\infty}
\newcommand{\ten}{\otimes}

\newtheorem{definition}{Definition}[chapter]
\newtheorem{proposition}{Proposition}[chapter]
\newtheorem{theorem}{Theorem}[chapter]
\newtheorem{lemma}{Lemma}[chapter]
\newtheorem{corollary}{Corollary}[chapter]
\newtheorem{remark}{Remark}[chapter]
\newtheorem{example}{Example}[chapter]

\newcommand{\por}{\kP_{0}(\kR)}

\newcommand{\pr}{\kP(\kR)}
\newcommand{\ph}{\kP(\lh)}

\newcommand{\poh}{\kP_{0}(\lh)}
\newcommand{\puh}{\kP_{1}(\lh)}

\newcommand{\pa}{\kP(\kaa)}

\newcommand{\pmm}{\kP(\mm)}
\newcommand{\lpa}{lin_{\kcc}\kP(\kaa)}
\newcommand{\orr}{\kO(\kR)}
\newcommand{\omm}{\kO(\mm)}

\newcommand{\qr}{\mathcal{Q}(\mathcal{R})}

\newcommand{\qpr}{\mathcal{Q}_{P}(\mathcal{R})}
\newcommand{\qpa}{\mathcal{Q}_{P}(\mathcal{A})}

\newcommand{\qpja}{\mathcal{Q}_{P_{j}}(\mathcal{A})}
\newcommand{\qpka}{\mathcal{Q}_{P_{k}}(\mathcal{A})}
\newcommand{\qh}{\mathcal{Q}(\mathcal{H})} 
\newcommand{\qa}{\mathcal{Q}(\mathcal{A})}

\newcommand{\qmm}{\mathcal{Q}(\mm)}

\newcommand{\ql}{\mathcal{Q}(\LL)}

\newcommand{\qal}{\mathcal{Q}_{a}(\LL)}

\newcommand{\lh}{\mathcal{L}(\mathcal{H})}

\newcommand{\all}{\forall}
\newcommand{\ex}{\exists}
\newcommand{\rr}{\kR}

\newcommand{\hr}{\kR_{sa}}
\newcommand{\el}{E_{\gl}}
\newcommand{\jl}{\kj_{\gl}}
\newcommand{\jm}{\kj_{\mu}}
\newcommand{\emm}{E_{\mu}}

\newcommand{\eal}{E^{A}_{\gl}}

\newcommand{\eamu}{E^{A}_{\mu}}

\newcommand{\epl}{E^{P}_{\gl}}

\newcommand{\ea}{E^{A}}

\newcommand{\ef}{E^{f}}
\newcommand{\ep}{E^{P}}

\newcommand{\we}{\wedge}
\newcommand{\We}{\bigwedge}
\newcommand{\Ve}{\bigvee}

\newcommand{\tto}{\mapsto}
\newcommand{\lra}{\Longrightarrow}
\newcommand{\llra}{\Longleftrightarrow}
\newcommand{\smm}{\setminus}
\newcommand{\dl}{\kD(\LL)}

\newcommand{\dr}{\kD(\rr)}
\newcommand{\dpr}{\kD_{P}(\rr)}
\newcommand{\dprr}{\kD_{pr}(\rr)}
\newcommand{\dal}{\kD_{a}(\LL)}
\newcommand{\dbl}{\kD_{b}(\LL)}

\newcommand{\irr}{\in \RR}

\newcommand{\lir}{\gl \in \RR}

\newcommand{\inn}{\in \NN}
\newcommand{\nin}{n \in \NN}

\newcommand{\urb}{\overset{-1}}

\newcommand{\pcx}{P_{\CC x}}
\newcommand{\pcen}{P_{\CC e_{n}}}

\begin{document}

\title{\Huge{Observables \\ II : Quantum Observables}}

\author{Hans F.\ de Groote\footnote{degroote@math.uni-frankfurt.de;
FB Mathematik, J.W.Goethe-Universität Frankfurt a.\ M.}}

\titlepage
\maketitle
\begin{abstract}
    In this work we discuss the notion of observable - both quantum
    and classical - from a new point of view. In classical mechanics, 
    an observable is represented as a function (measurable, continuous
    or smooth), whereas in (von Neumann's approach to) quantum
    physics, an observable is represented as a bonded selfadjoint
    operator on Hilbert space. We will show in the present part II and
    the forthcoming part III of this work that there is a common structure
    behind these two different concepts. If $\mathcal{R}$ is a von Neumann
    algebra, a selfadjoint element $A \in \mathcal{R}$ induces a continuous
    function $f_{A} : \mathcal{Q}(\mathcal{P(R)}) \to \mathbb{R}$ defined
    on the \emph{Stone spectrum} $\mathcal{Q}(\mathcal{P(R)})$
    (\cite{deg3}) of the lattice $\mathcal{P(R)}$ of projections in
    $\mathcal{R}$. $f_{A}$ is called the observable function corresponding
    to $A$. The aim of this part is to study observable functions and its
    various characterizations.
\end{abstract}

\begin{center}
    {\large Für Karin}
\end{center}

\tableofcontents

\chapter{Introduction}
\label{in}

\pagestyle{myheadings}
\markboth{Introduction}{Introduction}

\begin{quote}
    \emph{``Neue Blicke durch die alten Löcher''\\
    (Georg Christoph Lichtenberg, Aphorismen)}
\end{quote}
~\\
~\\
In von Neumann's axiomatic approach to quantum physics, an observable 
property of a quantum system is represented by a selfadjoint bounded
linear operator on a suitable complex Hilbert space. In classical
mechanics, however, an observable property is, depending on the
context, represented by a measurable, continuous or smooth function on
phase space. In this and in the next parts (\cite{deg4, deg5}) we will
show how to overcome this apparently fundamental difference. \\
~\\
In the previous part (\cite{deg3}), we have studied the Stone spectrum
$\ql$ of a lattice $\LL$. The elements of $\ql$ are the maximal dual
ideals in $\LL$. $\ql$ is equipped with a topology by the requirement 
that the sets
\[
    \qal := \{ \frb \in \ql \mid a \in \frb \} \quad (a \in \LL) 
\]
form a basis of this topology. Of course, this is a manifest
generalization of Stone's construction (\cite{birk}). If $\kS$ is a
presheaf on a complete lattice $\LL$, the sheafification of $\LL$
leads to the same construction of $\ql$ as base space of the etale
space of $\kS$. Moreover, as we have proved in \cite{deg3}, if $\LL$
is the lattice $\pr$ of projections in an abelian von Neumann algebra 
$\rr$, then $\qr$ is homeomorphic to the Gelfand spectrum of $\rr$. In
the same way we define a topology on the set $\dl$ of all dual ideals 
in $\LL$. The Stone spectrum $\ql$ is a dense subset of $\dl$, but
note that, except for trivial situations, the topology of $\dl$ is not
Hausdorff.\\
~\\
If $E = (\el)_{\lir}$ is a bounded spectral family in a complete lattice
$\LL$, we call the function $f_{E} : \dl \to \RR$, defined by
\[
    f_{E}(\kj) := \inf \{ \lir \mid \el \in \kj \},  
\]
the \emph{observable function corresponding to $E$.} The aim of this
part is to study observable functions and its various
characterizations, in particular in the case $\LL = \pr$ for a von Neumann
algebra $\rr$. In that case, bounded spectral families correspond to
selfadjoint elements of $\rr$. Hence we write $f_{A}$ for the observable
function corresponding to the spectral family $\ea$ of $\hr$. We prove
that $f_{A} : \qr \to \RR$ is a \emph{continuous} function whose range
is the spectrum of $A$. Moreover, we show that, if $\rr$ is abelian,
the mapping $A \tto f_{A}$ coincides with the Gelfand transformation.
Thus it is tempting to regard $\qr$ as a sort of phase space. This interesting
question will be discussed in part IV (\cite{deg5}). \\
We refer to the introduction of part I (\cite{deg3}) for a detailed
description of results of the present part. 

\chapter{Quantum Observables}
\label{QO}

Observables in \emph{quantum physics} are
selfadjoint operators of an appropriate Hilbert space $\kh$. Physically
meaningful is not the \emph{precise value} of an observable (which is an
inconsistent notion in quantum theory by the Kochen - Specker theorem
(\cite{doe})) but its \emph{expectation value} in a given state of the 
physical system. In quantum theory the expectation that the value of
the observable $A$ lies in the Borel set $\Delta \tm \RR$ when the
physical system is in the pure state $x \in \kh$ is given by
\[
    <E(\Delta)x, x> = \int_{\Delta}\gl d<E_{\gl}x, x>
\]    
where $E = (E_{\gl})_{\gl \in \RR}$ is the spectral resolution
of the selfadjoint operator $A$. So the essence
of an observable is its spectral family. Later on we will show how to
describe also classical observables by spectral families. Of course
one can object that one can easily perform algebraic operations on
operators and functions but it is an intricate problem to describe
these operations in the language of spectral families. From the
physical point of view however, the possibility of adding two given
observables to obtain a new one is merely a mathematical reflex: what 
is the meaning of the sum of the position and the momentum operator or
the sum of two different spin operators? \\ 
The aim of the following section is to show how the representation of
quantum observables as observable functions evolves from sheaf
theoretical considerations.

\section{Motivation: The Presheaf of Spectral Families}
\label{PSF}
\pagestyle{myheadings}
\markboth{Quantum Observables}{Motivation: The Presheaf of Spectral Families}
   
   We know that there is only the trivial sheaf on the quantum 
   lattice $\LL(\kh)$ of closed subspaces of the Hilbert space $\kh$. 
   But what about presheaves?
   
   An obvious example is the following one: For $U \in \LL(\kh)$ let 
   $\kS(U) := \kL(U)$ be the space of bounded linear operators $U \to 
   U$ and for $V \in \LL(\kh),\ V \tm U$, we define a ``restriction map''
   \[
   \rho_{V}^{U} : \kL(U) \to \kL(V)
   \]
   by
   \[ \rho_{V}^{U}(A) := P_{V}A\mid_{V}.
   \]
   Clearly these data give a presheaf on $\LL(\kh)$.\\
   ~\\
   This example looks somewhat artificial because the restriction maps 
   defined above do not coincide with the usual idea of restricting a 
   mapping from its domain to a smaller set. But we will see in part
   III, that it leads quite naturally to the notion of \emph{Positive 
   Operator Valued Measures}. The elements of the stalks of this
   presheaf, however, have a quantum mechanical interpretation.
   
   \begin{remark}\label{m1}
   Let $A \in \kL(U)$ and let $\frb_{\CC x} \in \kQ_{U}(\kh)$ be an 
   atomic quasipoint (in $\LL(U)$). Then the germ of $A$ in $\frb_{\CC 
   x}$ is given by $<Ax, x>$, where $x \in S^1(\kh) \cap \CC x$.
   \end{remark}
   Namely, if $A, B \in \kL(U)$, then $A \sim_{\frb_{\CC x}} B$ if and 
   only if $P_{\CC x}AP_{\CC x} = P_{\CC x}BP_{\CC x}$. Now if $x \in 
   S^1(\kh)$ then
   \[
   \forall \ z \in \kh : P_{\CC x}AP_{\CC x}z = <Ax, x><z, x>x.
   \]
   Hence $P_{\CC x}AP_{\CC x} = P_{\CC x}BP_{\CC x}$ if and only if 
   $<Ax, x> = <Bx, x>$.\\
   ~\\
   If $A$ is a selfadjoint operator and $x\in S^1(\kh)$ is in the
   domain of $A$, then $<Ax, x>$ is interpreted as the \emph{expectation
   value of the observable $A$ when the quantum mechanical system is in the
   pure state $\CC x$.}\\
   ~\\
   In order to obtain an example of a presheaf on $\LL(\kh)$ whose
   restriction maps are defined analogously to the ordinary restriction of
   functions, we shall reformulate the operation of restricting a 
   continuous function $f : U \to \RR$ to an open subset $V \tm U$ in 
   the language of lattice theory.\\
   ~\\
   Let $M$ and $N$ be regular Hausdorff spaces. A continuous mapping 
   \mbox{$f : M \to N$} induces a lattice homomorphism
   \begin{eqnarray*}
	\Phi_{f} : \kT(N) & \to & \kT(M)  \\
	W & \mapsto & \overset{-1}{f}(W)
   \end{eqnarray*}
   that is left continuous:
   \[
   \Phi_{f}(\bigcup_{i\in I}W_{i}) = \bigcup_{i\in I}\Phi_{f}(W_{i})
   \]
   for each family $(W_{i})_{i\in I}$ in $\kT(N)$. Conversely:
   
   \begin{theorem}\label{m2}
   Each left continuous lattice homomorphism $\Phi : \kT(N) \to \kT(M)$ 
   induces a unique continuous mapping $f : M \to N$ such that 
   $\Phi = \Phi_{f}$.
   \end{theorem}
   The proof is based on the observation that for any point $\frp$ 
   in $\kT(M)$ the inverse image $\overset{-1}{\Phi}(\frp)$ is a point 
   in $\kT(N)$. Because the points in $\kT(M)$ correspond to the 
   elements of $M$, this gives a mapping $f : M\to N$. It is then easy 
   to show that $f$ has the required properties.\\
   Now we can describe the restriction of a continuous mapping $f : M 
   \to N$ to an open subset $U$ of $M$ in the following way:
   
   \begin{proposition}\label{m3}
   Let $f : M \to N$ be a continuous mapping between regular Hausdorff 
   spaces, $\Phi_{f} : \kT(N) \to \kT(M)$ the left-continuous lattice 
   homomorphism induced by $f$, and $U$ an open subset of $M$. Then
   \begin{eqnarray*}
	\Phi_{f}^{U} : \kT(N) & \to & \kT(U)  \\
	W & \mapsto & \Phi_{f}(W) \cap U
    \end{eqnarray*}
    is a left-continuous lattice homomorphism and the corresponding 
    continuous mapping $U \to N$ is the restriction of $f$ to $U$.
    \end{proposition}
   \emph{Proof:} $\gF := \Phi_{f}^{U} : \kT(N) \to \kT(U)$ is a 
   left-continuous lattice homomorphism, since $\kT(M)$ is a
   completely distributive lattice. Let $\frp_{x} \tm \kT(U)$ be the
   point corresponding to $x \in U$. Then
   \[
       \urb{\gF}(\frp_{x}) = \{ V \in \kT(N) \ | \ x \in \gF_{f}(V)
       \cap U \} = \{ V \in \kT(N) \ | \ x \in \gF_{f}(V) \} =
       \frp_{f(x)},
   \]
   where $\frp_{f(x)}$ denotes the point in $\kT(N)$ that corresponds 
   to $f(x)$. Hence $\gF = \gF_{f_{|_{U}}}$. \ \ $\Box$ \\
    
    Let $\kh$ be a Hilbert space. The observables of a quantum 
    mechanical system are selfadjoint operators of $\kh$. Equivalently 
    we can think of observables as spectral families in the lattice
    $\ph$ of all orthogonal projections in $\lh$. Let $\kS(\kh)$ be
    the set of all spectral families in $\ph$. To begin with, we restrict
    our attention to those spectral families $E : \RR \to \ph$ that are
    \emph{bounded from above} :
    \[
    \exists \ \gl \in \RR \ : \ \el = I .
    \]
    We want to show that the set $\kS^{ub}(\kh)$ of all \emph{upper
    bounded} spectral families in $\ph$ induces canonically a presheaf
    $\kS^{ub}_{\kh}$ on $\ph$. We can perform the construction for an
    arbitrary orthomodular lattice. \\
    ~\\
    Let $\LL$ be a complete lattice and for $a \in \LL$ let 
    \[
        \LL_{a} := \{ b \in \LL \mid b ≤ a \}.
    \]
    $\LL_{a}$ is a complete orthomodular sublattice (in fact a
    principle ideal) of $\LL$ with maximal element $a$. We denote by
    $\kS^{ub}(a)$ the set of all spectral families $E : \RR \to \LL_{a}$
    that are bounded from above:
    \[
        \exists \ \gl \in \RR \ : \ \el = a.
    \]
    For $a, b \in \LL, \ a ≤ b$ we define a restriction mapping 
    \[
        \begin{array}{cccc}
            \gr^b_{a} : & \kS^{ub}(b) & \to & \kS^{ub}(a)  \\
             & E & \tto & E^{a}
        \end{array}
    \]  
    by
    \[
        \all \ \lir : E^{a}_{\gl} := \el \we a.
    \]
    Obviously,
    \[
        \kS^{ub}_{\LL} := (\kS^{ub}(a), \gr^b_{a})_{a ≤ b}
    \]
    is a presheaf on $\LL$. We call it the {\bf spectral presheaf on
    $\LL$}.
    
    \begin{remark}\label{m3a}
        If $\LL$ is a lattice of finite type (\cite{deg3}), the condition
        of upper boundedness is not necessary. If, in particular,
        $\LL$ is the projection lattice $\pr$ of a finite von Neumann 
        algebra, the restriction maps are defined for arbitrary
        spectral families. Theorem 3.1 in \cite{deg3} and its proof
        show that the converse is true too. 
    \end{remark}
    We can make the connection of restricting spectral familes to the 
    restriction of continuous real valued functions on a Hausdorff
    space even more transparent:\\
    Let $f : M \to \RR$ be a continuous function on a Hausdorff space 
    $M$. Then 
    \[
	\all \ \gl \irr : \ \el := int (\urb{f}(]-\∞, \gl ]))
    \]
    defines a spectral family $E : \RR \to \ktm$. (The natural
    guess for defining a spectral family corresponding to $f$ would be
    \[
	\gl \tto \urb{f}(]-\∞, \gl [).
    \]
    In general, this is only a \emph{pre-spectral family}: it
    satisfies all properties of a spectral family, except continuity
    from the right. This is cured by \emph{spectralization}, i.e. by
    the switch to
    \[
	\gl \tto \We_{\mu > \gl}\urb{f}(]-\∞, \mu [).
    \]
    But
    \[
    \We_{\mu > \gl}\urb{f}(]-\∞, \mu [) = int (\bigcap_{\mu > \gl}
    \urb{f}(]-\∞, \mu [)) = int (\urb{f}(]-\∞, \gl ])),
    \]
    which shows that our original definition is the natural one.)\\
    One can show that
    \[
	\all \ x \in M : \ f(x) = \inf \{ \gl \ | \ x \in \el \},
    \]
    so one can recover the function $f$ from its spectral family $E$. 
    Let $U \in \ktm, \ U \ne \emptyset$. Then 
    \[
	\el \cap U = int \{ x \in U \ | \ f(x) ≤ \gl \} =
	int(\urb{f_{|_{U}}}(]-\∞, \gl])),  
    \]
    hence $E^{U}$ is the spectral family of the usual restriction
    $f_{|_{U}} : U \to \RR$ of $f$ to $U$. We shall investigate the
    interplay between spectral families in $\ktm$ and continuous
    functions $f : M \to \RR$ extensively in the next part. \\
    ~\\
    Let $\kh$ be a Hilbert space. The following simple example shows
    that the restriction $E^{P}$ of a spectral family $E$ in $\ph$ that
    is not bounded from above may fail to be a spectral family in
    $\kP(P\kh)$.

    \begin{example}\label{m4}
    Let $\kh$ be a separable Hilbert space and $(e_{n})_{n \in \NN}$ 
    an orthonormal basis of $\kh$. Then
    \[
    \el := \bigvee_{n \leq \gl}\pcen \qquad (\gl \in \RR)
    \]
    defines a spectral family in $\LL(\kh)$. One can show that this 
    spectral family corresponds (up to some scaling) to the Hamilton 
    operator of the harmonic oscillator. Take $x \in S^1(\kh)$ such 
    that
    \[
    \forall \ n \in \NN \ : \ <x, e_{n}> \ne 0.
    \]
    This means that $\pcx \nleq \el$ for all $\gl \in \RR$ and hence
    \[
    E^{\pcx}_{\gl} = \el \we \pcx = 0
    \]
    for all $\gl \in \RR$. Therefore
    \begin{displaymath}
	\bigvee_{\gl \in \RR}E^{\pcx}_{\gl} = 0 \ne \pcx.
    \end{displaymath}
    \end{example}
    
    \begin{remark}\label{m5}
    Of course we can drop the requirement
    \begin{displaymath}
	\bigvee_{\gl \in \RR}\el = I
    \end{displaymath}
    in the definition of spectral families. Then we obtain the notion 
    of a {\bf generalized spectral family}. Operators that are given 
    by generalized spectral families are not necessarily densely 
    defined, but their domain of definition is only dense in the 
    closed subspace $\bigvee_{\gl \in \RR}\el \kh$ of $\kh$.
    \end{remark}
    Let us consider the restriction of a spectral family $E : \RR 
    \to \ph$ to $\kP(\pcx \kh)$ more closely. 
    If $\pcx ≤ \el$ for some $\gl \in \RR$, then the hermitian 
    operator corresponding to the spectral family
    \begin{displaymath}
	E^{\pcx} : \RR \to \kP(\pcx \kh)
    \end{displaymath}
    is a (real) scalar multiple $cI_{1}$ of the identity $I_{1} : \pcx 
    \kh \to \pcx \kh$. Now $\kP(\pcx \kh) = \{0, \pcx \}$, hence
    \begin{displaymath}
	E^{\pcx}_{\gl} =
	\begin{cases}
	0     &\text{for $\gl < c$}\\
	\pcx &\text{for $\gl \geq c$}
	\end{cases}
    \end{displaymath}
    and
    \begin{displaymath}
	c = \inf \{\gl \in \RR \mid \pcx ≤ \el \}.
    \end{displaymath}
    Using the convention
    \begin{displaymath}
	\inf \emptyset = \infty
    \end{displaymath}
    we obtain in this way a function on the projective Hilbert space 
    $\PP\kh$ with values in $\RR \cup \{\infty\}$,
    \begin{displaymath}
	f_{E} : \PP\kh \to \RR \cup \{\infty\},
    \end{displaymath}
    defined by
    \begin{displaymath}
	f_{E}(\CC x) := \inf \{\gl \in \RR \mid \pcx ≤ \el \}.
    \end{displaymath}
    Clearly, if $E$ is bounded from above then $f_{E}$ is bounded 
    from above, too. Moreover, $f_{E}$ is a bounded function if and 
    only if $E$ is a bounded spectral family, i.e. the corresponding 
    selfadjoint operator $A_{E}$ is bounded.
    
    The canonical topology on projective Hilbert space $\PP\kh$ is the 
    quotient topology defined by the projection
    \begin{eqnarray*}
	pr : \kh \setminus\{0\} & \to & \PP\kh  \\
	x & \mapsto & \CC x.
    \end{eqnarray*}
    This means that a subset $\mathcal{W} \tm \PP\kh$ is open if and 
    only if $\overset{-1}{pr}(\mathcal{W})$ is an open subset of $\kh 
    \setminus\{0\}$.\\
    The function $f_{\gs}$ has some remarkable properties:
    \begin{proposition}\label{m6}
    Let $E : \RR \to \ph$ be a spectral family and let
    \begin{displaymath}
	f_{E} : \PP\kh \to \RR \cup \{\infty\}
    \end{displaymath}
    be the function defined by
    \begin{displaymath}
	f_{E}(\CC x) := \inf \{\gl \in \RR \mid \pcx ≤ \el \}.
    \end{displaymath}
    Then
    \begin{enumerate}
        \item  [(1)] $f_{E}$ is lower semicontinuous on $\PP\kh$;
    
        \item  [(2)] if $\CC x, \CC y, \CC z$ are elements of $\PP\kh$ 
	such that $\CC z \tm \CC x + \CC y$, then
	      \begin{displaymath}
	f_{E}(\CC z) \leq \max (f_{E}(\CC x), f_{E}(\CC y));
	      \end{displaymath}
    
        \item  [(3)] $\overset{-1}{f_{E}}(\RR)$ is dense in $\PP\kh$.
    \end{enumerate}
    \end{proposition}
    Lower semicontinuity follows from
    \begin{displaymath}
	\overset{-1}{pr}(\overset{-1}{f_{E}}(]- \infty, \gl])) \cup 
	\{0\} = \el \kh;
    \end{displaymath}
    for then $\overset{-1}{f_{E}}(]-\infty, \gl])$ is closed in 
    $\PP\kh$ for all $\gl \in \RR$ and therefore $f_{E}$ is lower 
    semicontinuous. The two other properties are obvious from the 
    definitions.
    
    \begin{definition}\label{m7}
    A function $f : \PP\kh \to \RR \cup \{\infty\}$ is called an 
    {\bf observable function} if it has the following properties:
    \begin{enumerate}
        \item  [(1)] $f$ is lower semicontinuous; 
    
        \item  [(2)] if $\CC x, \CC y, \CC z$ are elements of $\PP\kh$ 
	such that $\CC z \tm \CC x + \CC y$ then
	\begin{displaymath}
	f(\CC z) \leq \max (f(\CC x), f(\CC y));
	\end{displaymath}
    
        \item  [(3)] $\overset{-1}{f}(\RR)$ is dense in $\PP\kh$.	
    \end{enumerate}
    \end{definition}
    The  point is that we can reconstruct spectral families 
    in $\ph$ from observable functions on $\PP\kh$:
    
    \begin{theorem}\label{m8}
    The mapping $E \mapsto f_{E}$ is a bijection from the set of 
    spectral families in $\ph$ onto the set of observable 
    functions on $\PP\kh$. This mapping is compatible with restrictions:
    \begin{displaymath}
	f_{E^{P}} = f_{E}|_{\PP P \kh}.
    \end{displaymath}
    Moreover, $E \in \kS^{ub}(\kh)$ if and only if 
    $\overset{-1}{f_{E}}(\RR) = \PP\kh$, and $E$ belongs to
    $\kS^{b}(\kh)$, the set of bounded spectral families in $\ph$, 
    if and only if $f_{E}$ is bounded.
    \end{theorem}
    \emph{Sketch of proof:} The construction of a spectral family from 
    an observable function $f$ is roughly as follows: for $\gl \in 
    \RR$ let
    \begin{displaymath}
	\el := \overset{-1}{pr}(\overset{-1}{f}(]-\infty, \gl])) 
	\cup \{0\}.
    \end{displaymath}
    Property $(1)$ assures that $\el$ is closed in $\kh$ and 
    property $(2)$ implies that $\el$ is a subspace of $\kh$. It 
    is not difficult to show that $E : \gl \mapsto \el$ is a 
    spectral family in $\ph$ and that
    \begin{displaymath}
	f_{E} = f
    \end{displaymath}
    holds. It follows from Baire's category theorem that $E \in 
    \kS^{ub}(\kh)$ if and only if $\overset{-1}{f_{E}}(\RR) = 
    \PP\kh$. \ $\Box$ \\
    ~\\
    If $A$ is the selfadjoint operator corresponding to the spectral 
    family $E$, then we also write $f_{A}$ instead of $f_{E}$.
    The {\bf spectrum} $sp(A)$ of a selfadjoint operator on $\kh$ is 
    given by the corresponding observable function $f_{A}$ in a 
    surprisingly simple manner:
    
    \begin{proposition}\label{m9}
    Let $A$ be a selfadjoint operator on $\kh$. Then
    \begin{displaymath}
	sp(A) = \overline{f_{A}(\overset{-1}{f_{A}}(\RR))},
    \end{displaymath}
    which simplifies to
    \begin{displaymath}
	sp(A) = \overline{f_{A}(\PP\kh)}
    \end{displaymath}
    if $A$ is bounded from above.
    \end{proposition}
    In the next section we will obtain a stronger result (theorem \ref{theo:
    3}) for bounded selfadjoint operators. \\
    ~\\
    $\kO(\kh)$ be the set of observable functions on $\PP\kh$, 
    $\kO^{ub}(\kh)$ the set of observable functions that are bounded 
    from above, and $\kO^{b}(\kh)$ the set of bounded observable 
    functions.\\
    Let $f \in \kO(\kh), \ \CC x \in \overset{-1}{f}(\RR) \quad 
    \text{and} \quad 
    \frb_{\CC x} \in \qh$ the atomic quasipoint defined by 
    $\CC x$. Let further $E$ be the spectral family corresponding 
    to $f$. Then
    \begin{eqnarray*}
	f(\CC x)   & = & \inf \{\gl \in \RR \mid \pcx ≤ \el \}  \\
	 & = & \inf \{\gl \in \RR \mid \el \in \frb_{\CC x} \}.
    \end{eqnarray*}
    Using this formulation, we can extend the definition of observable 
    functions to arbitrary quasipoints in $\ph$:
    
    \begin{definition}\label{m10}
    Let $f \in \kO(\kh)$ and let $E_{f} : \RR \to \ph$ be the 
    spectral family corresponding to $f$. The function
    \begin{displaymath}
	\hat{f} : \qh \to \RR \cup \{-\∞, \infty\},
    \end{displaymath}
    defined by
    \begin{displaymath}
	\hat{f}(\frb) := \inf \{\gl \in \RR \mid \el \in \frb \}, 
    \end{displaymath}
    is called the {\bf observable function on $\qh$ induced 
    by $f$.}
    \end{definition}
    Note that $\hat{f}(\frb) = -\∞$ if and only if $E$ is not bounded 
    from below and $\frb$ contains $\{ \el \mid \lir \}$. \\
    The observable function $\hat{f}$ induced by $f \in \kO^{b}(\kh)$ 
    can also be expressed directly in terms of $f$:
    
    \begin{proposition}\label{m11}
    Let $f$ be a bounded observable function. Then the observable 
    function $\hat{f}$ induced by $f$ is given by
    \begin{displaymath}
        \forall \ \frb \in \qh \ : \ \hat{f}(\frb) = \inf_{ P 
	\in \frb}\sup_{\pcx ≤ P}f(\CC x).
    \end{displaymath}
    \end{proposition}
    From now on we will denote the observable function $\qh 
    \to \RR \cup \{-\∞, \infty\}$ induced by $f \in \kO(\kh)$ also with 
    the letter $f$. \\
    ~\\
    Next we will show how observable functions $f : \qh \to \RR \cup
    \{-\∞\}$ can be used to assign a value to the germ $[E]_{\frb}$ of 
    a spectral family $E$ in the quasipoint $\frb \in \qh$. 
    We recall that spectral families $E \in \kS^{ub}(P)$ and $F 
    \in \kS^{ub}(Q)$ are equivalent at the quasipoint $\frb \in \kQ_{ 
    P \we Q}(\kh)$ if and only if there is an element $R \in \frb$ 
    such that $R ≤ P \we Q$ and $E^{R} = F^{R}$ holds.
    
    \begin{proposition}\label{m12}
    Let $E \in \kS^{ub}(P), \ F \in \kS^{ub}(Q)$ be spectral 
    families with corresponding observable functions $f_{E}$ and 
    $f_{F}$ respectively. If $E$ and $F$ are equivalent at 
    $\frb \in \kQ_{P \we Q}(\kh)$, then
    \begin{displaymath}
	f_{E}(\frb) = f_{F}(\frb)
    \end{displaymath}
    holds.
    \end{proposition}
    This follows directly from the observation that the definition of 
    equivalence at $\frb$ implies
    \begin{displaymath}
        \{\gl \in \RR \mid \el \in \frb \} = \{\gl \in \RR \mid 
	F_{\gl} \in \frb \}.
    \end{displaymath}
    The proposition shows that we obtain a mapping
    \[	v : \kE(\kS^{ub}_{\kh})  \to  \RR \cup \{-\∞ \}  \]
    defined by
    \begin{displaymath}
		v([E]_{\frb}) = f_{E}(\frb)
    \end{displaymath}
    on the etale space $\kE(\kS^{ub}_{\kh})$. $v([E]_{\frb})$ is called 
    the {\bf value of the germ $[E]_{\frb}$.}

\section{Basic Properties}
\label{sec: BQOb}
\pagestyle{myheadings}
\markboth{Quantum Observables}{Basic Properties}

Let $\kh$ be a Hilbert space and let $A$ be a selfadjoint (not
necessarily bounded) operator of $\kh$. Our basic definition is
\begin{definition}\label{def: 1}
    Let $\ea = (\eal)_{\gl \in \RR}$ be the spectral family of $A$.
    The function 
    \[
        f_{A} : \qh \to \RR \cup \{ - \∞, \∞ \},
    \]
    defined by 
    \[
        f_{A}(\frb) := \inf \{ \gl \in \RR | \ \eal \in \frb \},
    \]
    is called the {\bf observable function corresponding to $A$.}
\end{definition}

\begin{remark}\label{rem: 2}
    The value $\∞$ occurs for $f_{A}$ if and only if $\eal \neq I$ for
    all $\gl \in \RR$, i.e. if $\ea$ is not bounded from above.
    Analogously the value $- \∞$ occurs if and only if $\ea$ is not
    bounded from below. $A$ is bounded if and only if the observable 
    function $f_{A}$ is bounded. Note that this is already the case 
    if $f_{A}$ is real valued.\\
   \end{remark}

In the following let $\rr$ be a von Neumann algebra considered, as a 
subalgebra of $\lh$ for some Hilbert space $\kh$, such that the unit 
element of $\rr$ is the identity operator $I = id_{\kh} \in \lh$.
$\hr$ denotes the set of selfadjoint elements of $\rr$ and $\pr$ the
lattice of projections in $\rr$. Let $\qr$ be the \emph{Stone spectrum
of $\rr$}, i.e. the Stone spectrum of the complete lattice $\pr$.
If $A \in \hr$, we denote by $sp(A)$ the \emph{spectrum of $A$}
and by $\ea$ the \emph{spectral family of $A$}\footnote{We do not
consider restrictions of spectral families in this and the remaining
sections of this chapter, so there is no danger to confuse $\ea$ with a
restriction of some spectral family $E$.}. \\
~\\
In the following two sections we will generalize the results of the
foregoing section to arbitrary von Neumann algebras.

\begin{theorem}\label{theo: 3}
    Let $A \in \hr$ and let $f_{A} : \qr \to \RR$ be the observable
    function corresponding to $A$. Then 
    \[
	im f_{A} = sp(A).
    \] 
\end{theorem}
\emph{Proof:} The spectrum $sp(A)$ of $A$ consists of all $\gl \in
\RR$ such that the spectral family $\ea$ of $A$ is non-constant on
every neighbourhood of $\gl$. Assume that $\gl_{0} \in im f_{A}$, but 
$\gl_{0} \notin sp(A)$. Then there is some $\eps > 0$ such that
\[
    \all \ \gl \in ]\gl_{0} - \eps, \gl_{0} + \eps [ : \ \eal =
    \ea_{\gl_{0}}.
\]
Therefore, if $\frb \in \qr$ is in the inverse image of $\gl_{0}$ by
$f_{A}$, then $f_{A}(\frb) ≤ \gl_{0} - \eps$, a contradiction. Thus
$f_{A} \tm sp(A)$. \\
Conversely let $\gl_{0} \in sp(A)$. There are two (non-excluding)
possibilities:
\begin{itemize}
    \item [(i)]  There is a decreasing sequence $(\gl_{n})_{n \in \NN}$ such
    that $\gl_{0} = \lim_{n \to \∞}\gl_{n}$ and $\ea_{\gl_{n + 1}} <
    \ea_{\gl_{n}}$ for all $n$.

    \item [(ii)]  $\eal < \ea_{\gl_{0}}$ for all $\gl < \gl_{0}$.
\end{itemize}
If the first possibility occurs, we take a quasipoint $\frb \in \qr$
that contains all $\ea_{\mu} - \ea_{\gl_{0}}$ for $\mu > \gl_{0}$.
Then $\eal \in \frb$ for all $\gl > \gl_{0}$, but $\ea_{\gl_{0}} \notin
\frb$. Hence $f_{A}(\frb) = \gl_{0}$. If the second possibility occurs,
we take a quasipoint $\frb$ that contains all $\ea_{\gl_{0}} - \eal$
for $\gl < \gl_{0}$. Then $\ea_{\gl_{0}} \in \frb$ but, $\eal \notin
\frb$ for $\gl < \gl_{0}$. Hence $f_{A}(\frb) = \gl_{0}$. \ \ $\Box$ 

\begin{remark}\label{rem: 4}
    The foregoing proof shows that the infimum in the definition of
    observable functions can, in general, not be replaced by a minimum.
\end{remark}

\begin{example}\label{ex: 5}
    The observable function of a projection $P \in \pr$ is given by
    \[
	f_{P} = 1 - \chi_{\kQ_{I - P}(\rr)},
    \]
    where $\chi_{\kQ_{I - P}(\rr)}$ denotes the characteristic
    function of the open closed set $\kQ_{I - P}(\rr)$. Hence $f_{P}$ 
    is a continuous function.
\end{example}
\emph{Proof:} The spectral family $\ep$ of $P$ is
\[
    \epl =
    \begin{cases}
	0      &  \text{for} \ \  \gl < 0   \\
	I - P  &  \text{for} \ \  0 ≤ \gl < 1  \\
	I       &  \text{for} \ \  1 ≤ \gl .
    \end{cases}
\]
The assertion follows now directly from the definition of observable
functions. \ \ $\Box$ \\

This example can be generalized easily.\\
Using the fact that for $a \in \RR$ and $A \in \hr$
\[
    E^{A + aI} = \ea \circ T_{a}
\]
holds, where $T_{a}$ denotes the translation $\gl \tto \gl - a$, we
obtain 

\begin{lemma}\label{lem: 6}
    If $A \in \hr$ and $a \in \RR$, then $f_{A + aI} = a + f_{A}$.
\end{lemma}
\emph{Proof:} For all $\frb \in \qr$ we have
\begin{eqnarray*}
    f_{A + aI}(\frb) & = & \inf \{ \gl | \ E^{A + aI}_{\gl} \in \frb \}  \\
     & = & \inf \{ \gl | \ \ea_{\gl - a} \in \frb \}  \\
     & = & \inf \{ a + \gl - a | \ \ea_{\gl - a} \in \frb \}  \\
     & = & a + \inf \{ \gl | \ \eal \in \frb \}   \\
     & = & a + f_{A}(\frb). \ \ \Box 
\end{eqnarray*}
\\
Consider pairwise orthogonal projections $P_{1}, \ldots , P_{n} \in
\rr$, nonzero real numbers $a_{1} < \cdots < a_{n}$ and let $A :=
\sum_{j = 1}^{n}a_{j}P_{j}, \ P := \sum_{j = 1}^{n}P_{j}.$ Then for
all $a \in \RR$
\begin{eqnarray*}
    A - aI & = & A - aP_{1} - \cdots - aP_{n} - a(I - P)  \\
     & = & (a_{1} - a)P_{1} + \cdots + (a_{n} - a)P_{n} + (-a)(I - P).
\end{eqnarray*}
Choose $a > 0$ such that $a_{j} - a < 0$ for all $j = 1, \ldots , n$. 
If there is some $j_{0}$ such that $a_{j_{0}} < 0 < a_{j_{0} + 1}$
then
\[
    a_{1} - a < \cdots < a_{j_{0}} - a < -a < a_{j_{0} + 1} - a < \cdots
    < a_{n} - a < 0.
\]
( The cases when the $a_{k}$ are all positive or all negative are
handled analogously.) For $k = 1, \ldots , n + 1$ set
\[
    b_{k} :=
    \begin{cases}
       a_{k} - a   &   \text{for} \quad   k ≤ j_{0}  \\
       -a             &   \text{for} \quad  k = j_{0} + 1   \\
       a_{k - 1} - a  &  \text{for} \quad  k ≥ j_{0} + 2
    \end{cases}
\]
and
\[
    Q_{k} :=
    \begin{cases}
	P_{k}     &   \text{for} \quad  k ≤ j_{0}  \\
	I - P       &   \text{for} \quad  k = j_{0} + 1  \\
	P_{k - 1}   &   \text{for} \quad  k ≥ j_{0} + 2.
     \end{cases}
\]
$(Q_{1}, \ldots , Q_{n + 1})$ is an orthogonal decomposition of $I$
and therefore the spectral family of $A - aI = \sum_{k = 1}^{n +
1}b_{k}Q_{k}$is given by
\[
    E^{A - aI}_{\gl} =
    \begin{cases}
	0          &   \text{for} \quad \gl < b_{1}  \\
	Q_{1} + \cdots + Q_{k}   &   \text{for} \quad  b_{k} ≤ \gl <
	b_{k + 1}   \\
	I            &   \text{for} \quad  \gl ≥ b_{n + 1}.
    \end{cases}    
\]
From the definition of observable functions we obtain
\[
    f_{A - aI}(\frb) = b_{k} \quad \text{for} \quad Q_{1} + \cdots +
    Q_{k} \in \frb, \ \ Q_{1} + \cdots + Q_{k - 1} \notin \frb.
\]
Therefore, setting $Q_{0} := 0$, we get
\[
    f_{A - aI} = \sum_{k = 1}^{n + 1}b_{k}\chi_{\kQ_{Q_{1} + \cdots 
    + Q_{k}}(\rr) \setminus \kQ_{Q_{1} + \cdots + Q_{k - 1}}(\rr)}.
\]
Combining this result with lemma \ref{lem: 6} gives a proof of

\begin{proposition}\label{prop: 7}
    Let $P_{1}, P_{2}, \ldots , P_{n} \in \pr$ be pairwise orthogonal 
    projections and $A := \sum_{k = 1}^{n}a_{k}P_{k}$ with real
    coefficients $a_{1}, a_{2}, \ldots , a_{n}$. Then, setting $P_{0} 
    := 0$,
    \[
        f_{A} = \sum_{k = 1}^{n}a_{k}\chi_{\kQ_{P_{1} + \cdots + P_{k}}(\rr)
	\setminus \kQ_{P_{1} + \cdots + P_{k - 1}}(\rr)}.
    \]                       
    Consequently, $f_{A}$ is continuous.
\end{proposition}

\begin{corollary}\label{cor: 8}
    If $\rr$ is abelian and $A = \sum_{k = 1}^{n}a_{k}P_{k}$ as in
    proposition \ref{prop: 7}, then
    \[
        f_{A} = \sum_{k = 1}^{n}a_{k} \chi_{\kQ_{P_{k}}(\rr)}.
    \]
\end{corollary}
\emph{Proof:} If $\rr$ is abelian then the projection lattice $\pr$ is
distributive and therefore 
\[
    \kQ_{P_{1} + \cdots + P_{k}}(\rr) = \bigcup_{j =
    1}^{k}\kQ_{P_{j}}(\rr). \ \ \Box
\]

\begin{theorem}\label{theo: 9}
    Let $A \in \hr$. Then the observable function $f_{A} : \qr \to
    \RR$ is continuous.
\end{theorem}
\emph{Proof:} We know from theorem \ref{theo: 3} that the image of
$f_{A}$ is the spectrum $sp(A)$ of $A$. Let $a := min (sp(A))$, $b :=
max (sp(A))$ and let $\eps > 0$. Choose $\gl_{0} \in ]a - \eps, a[, \
\gl_{n} \in ]b, b + \eps[, \ \gl_{1}, \ldots , \gl_{n - 1} \in [a, b]$
such that $\gl_{k - 1} < \gl_{k}$ and $\gl_{k} - \gl_{k - 1} < \eps$
for $k = 1, \ldots , n$. Moreover let $\gl^{*}_{k} \in ]\gl_{k - 1},
\gl_{k}[$ for $k = 1, \ldots , n$ and
\[
    A_{\eps} := \sum_{k = 1}^{n}\gl^{*}_{k}(\ea_{\gl_{k}} -
    \ea_{\gl_{k - 1}}) := \sum_{k = 1}^{n}\gl^{*}_{k}P_{k}
\] 
with $P_{k} := \ea_{\gl_{k}} - \ea_{\gl_{k - 1}}$. Then by proposition
\ref{prop: 7}
\[
    f_{A_{\eps}} = \sum_{k = 1}^{n}\gl^{*}_{k}\chi_{\kQ_{P_{1} + \cdots
    + P_{k}}(\rr) \setminus \kQ_{P_{1} + \cdots + P_{k - 1}}(\rr)} =
    \sum_{k = 1}^{n}\gl^{*}_{k}\chi_{\kQ_{\ea_{\gl_{k}}}(\rr) \setminus
    \kQ_{\ea_{\gl_{k - 1}}}(\rr)}. 
\]
Let $\frb \in \qr$. Then $\frb \in \kQ_{\ea_{\gl_{k}}}(\rr) \setminus
    \kQ_{\ea_{\gl_{k - 1}}}(\rr)$ for exactly one $k$. Hence
    \[
	f_{A_{\eps}}(\frb) = \gl^{*}_{k}
    \]
    and
    \[
	f_{A}(\frb) = \inf \{ \gl | \ \eal \in \frb \} \in [\gl_{k -
	1}, \gl_{k}].
    \]
This implies
\[
    \mid f_{A}(\frb) - f_{A_{\eps}}(\frb) \mid < \eps 
\]    
and, $\frb$ being arbitrary,
\[
    \mid f_{A} - f_{A_{\eps}} \mid_{\∞} ≤ \eps.
\]
Hence $f_{A}$ is continuous. \ \ $\Box$ \\
    
\begin{definition}\label{def: 10}
    Let $\rr$ be a von Neumann algebra. Then we denote by $\orr$
    the set of observable functions $\qr \to \RR$.
\end{definition}

By the foregoing result, $\orr$ is a subset of $C_{b}(\qr, \RR)$, the
algebra of all bounded continuous functions $\qr \to \RR$. $\orr$
separates the points of $\qr$ because the observable function of a
projection $P$ is $f_{P} = 1 - \chi_{\kQ_{I - P}(\rr)}$. Moreover, it
contains the constant functions (by lemma \ref{lem: 6}). In general,
however, it is not an algebra and not even a vector space (with respect
to the pointwise defined algebraic operations).

\begin{theorem}\label{theo: 11}
    Let $\rr$ be a von Neumann algebra and let $\orr$ be the set of
    observable functions on $\qr$. Then 
    \[
	\orr = C_{b}(\qr, \RR)
    \]
    if and only if $\rr$ is {\bf abelian}.
\end{theorem}

We will prove here only one half of this theorem leaving the other
one until we have proved an abstract characterization of observable
functions.
\begin{proposition}\label{prop: 12}
    Let $\rr$ be a von Neumann algebra such that for all $P \in \pr$
    the characteristic function $\chi_{\kQ_{P}(\rr)}$ of the open
    closed set $\kQ_{P}(\rr)$ is an observable function. Then $\rr$ is
    abelian.
\end{proposition}
\emph{Proof:} Let $P_{0} \in \pr$. By assumption
$\chi_{\kQ_{P_{0}}(\rr)}$ is the observable function $f$ of an element
$A \in \hr$. Then $sp(A) = im f \tm \{0, 1\}$ and therefore $A$ is a
projection $P$ in $\rr$. Hence
\begin{equation}
    \chi_{\kQ_{P_{0}}(\rr)} = 1 - \chi_{\kQ_{I - P}(\rr)}.
    \label{eq: 1}
\end{equation}
Let $ \frb \in \kQ_{P}(\rr)$. Then $\chi_{\kQ_{P_{0}}(\rr)}(\frb) = 1$
and therefore $P_{0} \in \frb$. Hence we have shown
\begin{equation}
    \all \ \frb \in \qr : \ (P \in \frb \quad \Longrightarrow \quad
    P_{0} \in \frb).
    \label{eq: 2}
\end{equation}
We show that this implies
\begin{equation}
    P = P_{0}.
    \label{eq: 3}
\end{equation}
This is equivalent to
\begin{equation}
    1 = \chi_{\kQ_{P}(\rr)} + \chi_{\kQ_{I - P}(\rr)}
    \label{eq: 4}
\end{equation}
i.e. to
\begin{equation}
    \qr = \kQ_{P}(\rr) \cup \kQ_{I - P}(\rr).
    \label{eq: 5}
\end{equation}
By proposition 3.5 in part I (\cite{deg3}), this property is
equivalent to the distributivity of the lattice $\pr$, i.e. to the
commutativity of $\rr$.\\
Assume that \ref{eq: 3} does not hold, i.e. $P_{0} \we P < P$ or
$P < P_{0}$. If $P_{0} \we P < P$,
take some $\frb \in \qr$ with $P - P_{0} \we P \in \frb$. Then $P \in 
\frb$, hence also $P_{0} \in \frb$ (by \ref{eq: 2}) and therefore
$P_{0} \we P \in \frb$, a contradiction. This shows $P_{0} \we P = P$ 
i.e. $P ≤ P_{0}$. Assume that $P < P_{0}$. Let $\frb \in \qr$ such
that $P_{0} - P \in \frb$. Then $P_{0} \in \frb$ and therefore \ref{eq: 
1} implies $I - P \notin \frb$. But then
\[
    P_{0} - P = P_{0}(I - P) = P_{0} \we (I - P) \notin \frb,
\] 
a contradiction again. Hence \ref{eq: 3} holds. \ \ $\Box$\\
~\\
The foregoing proposition can be reformulated in the following way.\\
If $E = (\el)_{\lir}$ is a bounded (right-continuous) spectral family,
a natural candidate for an orthocomplement of $E$ in the lattice of
bounded spectral families (\cite{deg2}) would be
\[
    \Tilde{E} : \gl \tto I - E_{- \gl}.
\] 
But $\Tilde{E}$ is, continuous from the left and, in general, not from 
the right. Spectralization of $\Tilde{E}$ gives the (right-continuous)
spectral family 
\begin{eqnarray*}
    (\neg E)_{\gl} & := & \We_{\mu > \gl}(I - E_{- \mu})  \\
     & = & I - \Ve_{\mu < - \gl}\emm  \\
     & = & I - E_{-\gl -}.
\end{eqnarray*}
A routine calculation shows (\cite{deg2}) that, if $A \in \hr$ is the 
operator corresponding to $E$, then the  operator corresponding to
$\neg E$ is $- A$. Therefore, we obtain for the negative of the observable
function of $- A$:
\begin{eqnarray*}
    - f_{- A}(\frb) & = & - \inf \{ \lir \mid I - E_{- \gl -} \in \frb
 \}   \\
     & = & - \inf \{ - \gl \irr \mid I - E_{\gl -} \in \frb \}  \\
     & = & \sup \{ \lir \mid I - E_{\gl -} \in \frb \}  \\
     & = & \sup \{ \lir \mid I - E_{\gl} \in \frb \}.
\end{eqnarray*}
In particular, for a projection $P$ we get
\[
    - f_{- P} = \chi_{\qpr}.
\] 
 
\begin{corollary}\label{12a}
    A von Neumann algebra $\rr$ is abelian if and only if
    \[
         \orr = - \orr.
    \]
\end{corollary}
\emph{Proof:} If $\rr$ is abelian and $\frb \in \qr$, then $P \in
\frb$ or $I - P \in \frb$ for all $P \in \pr$. Hence
\[
    - f_{- A}(\frb) = \sup \{ \lir \mid I - \eal \in \frb \} = \inf
    \{ \lir \mid \eal \in \frb \} = f_{A}(\frb) 
\] 
for all $A \in \hr$ and all $\frb \in \qr$. \ \ $\Box$

\begin{remark}\label{12b}
    The functions
    \[
       \begin{array}{cccc}
           g_{A} : & \qr & \to & \RR  \\
            & \frb & \tto & \sup \{ \lir \mid I - \eal \in \frb \}
       \end{array}
    \]
    were introduced by Döring (\cite{doe1b}). His motivation came from
    the following observation. Let $A \in \lh_{sa}$ and $x \in \kh \smm
    \{0\}$. Then, by the spectral theorem, we have
    \[
        < A x, x > = \int_{- |A|}^{|A|}\gl d< \eal x, x>.
    \] 
    It is obvious that
    \[
        < \eal x, x > =
	\begin{cases}
	    < x, x >    & \ \text{for} \ \gl > f_{A}(\frb_{\CC x}) \\
	    0              & \ \text{for} \ \gl < g_{A}(\frb_{\CC x}), 
	\end{cases}
    \]
    where $\frb_{\CC x}$ is the atomic quasipoint defined by $\pcx$.
    Hence
    \[
        < A x, x > = \int_{g_{A}(\frb_{\CC x})}^{f_{A}(\frb_{\CC
	x})}\gl d< \eal x, x >.
    \]
    Döring called the function $g_{A} : \qr \to \RR$ the {\bf
    antonymous function} induced by $A \in \hr$. Because of $g_{A} = -
    f_{- A}$ for all $A \in \hr$, the set of antonymous functions is
    simply $- \orr$. Therefore, we prefer the name {\bf mirrored
    observable function} instead of the pretentious ``antonymous function''.
    Of course, the properties of mirrored observable functions are quite 
    analogous to those of observable functions. For example, we have
    \[
        im (- f_{- A}) = - im f_{- A} = - sp(- A) = sp(A).
    \]
    Similarly, it is quite easy to translate the whole discussion of
    the next sections to the mirrored case. Observable and mirrored
    observable functions are two sides of the same coin. Their
    symmetric rôle will become more clear in part IV (\cite{deg5}).
\end{remark}
\pagebreak

\section{Abstract Characterization of Observable Functions}
\label{BQObb}
\pagestyle{myheadings}
\markboth{Quantum Observables}{Abstract Characterization of Observable Functions}

We will prove some properties of observable functions of a von Neumann
algebra $\rr$ which in turn will serve as axioms for an abstract
notion of observable function.

\begin{definition}\label{def: 13}
    Let $\LL$ be a complete lattice (with minimal element $0$ and
    maximal element $1$). A nonempty subset $\kj \tm \LL$ is called a 
    {\bf dual ideal} if it has the following properties:
    \begin{enumerate}
        \item  [(i)] $0 \notin \kj$,
    
        \item  [(ii)] $a, b \in \kj \ \lra \ a \we b \in \kj$,
    
        \item  [(iii)] if $a \in \kj$ and $a ≤ b$, then $b \in \kj$.
    \end{enumerate}
    If $a \in \LL \smm \{0\}$, the dual ideal
    \[
        H_{a} := \{ b \in \LL | \ b ≥ a \}
    \]
    is called the {\bf principal dual ideal generated by $a$}.
    We denote by $\dl$ the set of dual ideals of $\LL$. For $a \in \LL$ let
    \[
        \dal := \{ \kj \in \dl | \ a \in \kj \}.
    \]
\end{definition}
As mentioned earlier, a maximal dual ideal is nothing but a quasipoint 
of $\LL$. We collect some obvious properties of the sets $\dal$ in
the following
\begin{remark}\label{rem: 14}
    For all $a, b \in \LL$ the following properties hold:
    \begin{enumerate}
        \item  [(i)] $a ≤ b \ \lra \ \dal \tm \dbl$, 
    
        \item  [(ii)] $ \kD_{a \we b}(\LL) = \dal \cap \dbl$,
    
        \item  [(iii)] $\dal \cup \dbl \tm \kD_{a \vee b}(\LL)$,
    
        \item  [(iv)] $\kD_{0}(\LL) = \lm, \ \kD_{1}(\LL) = \dl$.
    \end{enumerate}
    These properties show in particular that $\{ \dal | \ a \in \LL
    \}$ is a basis of a topology on $\dl$. The Stone spectrum $\ql$ is
    dense in $\dl$ with respect to this topology.
\end{remark}
Note that $\dl$ is in general {\bf not a Hausdorff space}: let $b, c
\in \LL, b < c$. Then $ H_{c} \in \dal \ \llra \ c ≤ a$, hence $H_{b} 
\in \dal$ and, therefore, $H_{b}$ and $H_{c}$ cannot be separated. We
return to the case that $\LL$ is the projection lattice $\pr$ of a von
Neumann algebra $\rr$, although most of our considerations also hold
for an arbitrary orthocomplemented complete lattice.

\begin{lemma}\label{lem: 15}
    $\all \ P \in \pr : \ H_{P} = \bigcap_{P \in \frb}\frb \ \  (=: \bigcap
    \qpr)$.
\end{lemma}
\emph{Proof:} If $P \in \frb$ then clearly $H_{P} \tm \frb$. Hence
$H_{P} \tm \bigcap \qpr$. Conversely, assume that $Q \in \bigcap \qpr
\smm H_{P}$. Then $P \we Q < P$ and so there is some quasipoint
$\frb$ which contains $P - P \we Q$. But then $P \in \frb$ and
therefore also $P \we Q \in \frb$, a contradiction. \ \ $\Box$ \\
~\\
Let $A \in \hr$ with corresponding spectral family $\ea$ and observable
function $f_{A}$. We extend $f_{A}$ to a function $\dr \to \RR$ on the
space $\dr$ of dual ideals of $\pr$ (and we denote this extension again
by $f_{A}$) in a natural manner:
\[
    \all \ \kj \in \dr : \ f_{A}(\kj) := \inf \{ \gl | \ \eal \in \kj
    \}.
\]
\begin{proposition}\label{prop: 16}
    Let $(\kj_{j})_{j \in J}$ be a family in $\dr$. Then
    \[
        f_{A}(\bigcap_{j \in J}\kj_{j}) = \sup_{j \in J}f_{A}(\kj_{j}).
    \]
\end{proposition}
\emph{Proof:} $\kj := \bigcap_{j \in J}\kj_{j}$ is a dual ideal that
is contained in $\kj_{j}$ for all $j \in J$. Hence 
\[
    f_{A}(\kj_{j}) = \inf \{ \gl | \ \eal \in \kj_{j} \} ≤ \inf \{ \gl | \ \eal \in \kj \}
    = f_{A}(\kj)
\]
and therefore 
\[
    \sup_{j} f_{A}(\kj_{j}) ≤ f_{A}(\kj).
\]
Let $\eps > 0 $ and choose $\gl$ such that
\[
    f_{A}(\kj) - \eps < \gl < f_{A}(\kj).
\]
Then $\eal \notin \kj$, so there is some $j_{0}$ such that $\eal
\notin \kj_{j_{0}}$. This means $f_{A}(\kj_{j_{0}}) ≥ \gl$ and
therefore we obtain
\[
    f_{A}(\kj) - \eps < f_{A}(\kj_{j_{0}}) ≤ \sup_{j} f_{A}(\kj_{j}).
\] 
As $\eps > 0$ was arbitrary, we conclude
\[
    f_{A}(\kj) ≤ \sup_{j} f_{A}(\kj_{j}).  \ \ \Box 
\]    
~\\
On a principal dual ideal $H_{P}$ we simply have 
\[
    f_{A}(H_{P}) = \inf \{ \gl | \ \eal ≥ P \} = min \{ \gl | \ \eal ≥ P \} 
\]    
because $\ea$ is continuous from the right.\\
~\\
The following characterization of eigenvalues of selfadjoint operators
is an application of the foregoing results. It is a generalization of 
5.7.22 in \cite{kr3} for the selfadjoint case. In particular, it gives
a further interesting characterization of finite von Neumann algebras.

\begin{proposition}\label{16a}
    Let $A \in \hr$ and $\gl \in sp(A)$. If $\gl$ is an eigenvalue, the
    interior of $\urb{f_{A}}(\gl)$ is a nonvoid subset of $\qr$.\\
    The converse holds if and only if $\rr$ is a finite von Neumann
    algebra.
\end{proposition}
\emph{Proof:} $\gl$ is an eigenvalue of $A$ if and only if $\eal -
\ea_{\gl -} > 0$. Take $\frb \in \kQ_{\eal - \ea_{\gl -}}(\rr)$. Then 
$\eal \in \frb$, hence $f_{A}(\frb) ≤ \gl$. If $f_{A}(\frb) < \gl$,
then $\eamu \in \frb$ for some $\mu < \gl$. But this implies $\ea_{\gl
-} \in \frb$, a contradiction. Hence $f_{A}(\frb) = \gl$ and,
therefore, $\kQ_{\eal - \ea_{\gl -}}(\rr) \tm \urb{f_{A}}(\gl)$. \\
Let $\rr$ be finite and let $\gl \in sp(A)$ such that $int
\urb{f_{A}}(\gl) \ne \emptyset$. Then there is some $P \in \por$ with 
$\qpr \tm \urb{f_{A}}(\gl)$. Since $H_{P} = \bigcap \qpr$ by lemma
\ref{lem: 15}, we obtain
\[
    f_{A}(H_{P}) = \sup_{\frb \in \qpr}f_{A}(\frb) = \gl.
\]
This implies 
\[
    \eal ≥ P.
\]
But then
\[
    \all \ \mu < \gl : \ \eamu \we P = 0,
\]
because, if $\eamu \we P \ne 0$ for some $\mu < \gl$, there is $\frb
\in \qpr$ with $\eamu \we P \in \frb$, hence $\eamu \in \frb$ and
therefore $f_{A}(\frb) ≤ \mu < \gl$, a contradiction. Now assume that 
$\eal = \ea_{\gl -}$. From the finiteness of $\rr$ we conclude 
\[
    0 = \Ve_{\mu < \gl}(P \we \eamu) = P \we \Ve_{\mu < \gl}\eamu = P 
    \we \eal = P,
\]
a contradiction. \\
If $\rr$ is not finite, we have to present an operator $A \in \hr$
that has a spectral value $\gl \in sp(A)$ such that $int
\urb{f_{A}}(\gl) \ne \emptyset$ and $\gl$ is not an eigenvalue. Assume
that $\rr$ is not finite. Then $\rr$ contains a direct summand of the form
$\mm \bar{\ten} \kL(\kh_{0})$, where $\mm \tm \kL(\kK)$ is a suitable
von Neumann algebra and $\kh_{0}$ a separable Hilbert space of infinite
dimension (see the proof of theorem 3.1 in \cite{deg3}). So it
suffices to find an example in $\mm \bar{\ten} \kL(\kh_{0})$. \\
Let $(e_{n})_{\nin}$ be an orthonormal basis of $\kh_{0}$ and for
$\lir$ let
\[
    \el :=
    \begin{cases}
	0          & \ \text{for} \ \gl < 0  \\
	\sum_{n = 1}^{k}\pcen    & \ \text{for} \ 1 - \frac{1}{k} ≤
	\gl < 1 - \frac{1}{k + 1} \ \text{and} \ k \inn  \\
	I           & \ \text{for} \  \gl ≥ 1.
    \end{cases}
\]
$(\el)_{\lir}$ is a bounded spectral family and therefore $(I_{\mm}
\ten \el)_{\lir}$ is a spectral family in $\mm \bar{\ten}
\kL(\kh_{0})$. $1$ belongs to the spectrum of the corresponding
selfadjoint operator $I_{\mm} \ten A \in \mm \bar{\ten} \kL(\kh_{0})$ 
but, by construction, $1$ is not an eigenvalue of $I_{\mm} \ten A$. We
have 
\[
    \rr = \mm \bar{\ten} \kL(\kh_{0}) \ \oplus \ \kS
\]
with a suitable von Neumann subalgebra $\kS$ of $\rr$. It is then easy
to see that a quasipoint in $\pr$ is either of the form $\frb \oplus
\kP(\kS)$, with $\frb \in \kQ(\mm \bar{\ten} \kL(\kh_{0}))$, or it is
of the form $\kP(\mm \bar{\ten} \kL(\kh_{0})) \oplus \frc$ with
$\frc \in \kQ(\kS)$. Now let $x := \sum_{n = 1}^{\∞}\frac{1}{n}e_{n}$.
Then the quasipoints that contain $(I_{\mm} \ten \pcx, 0)$ are of the 
form $\frb \oplus \kP(\kS)$. Thus we can restrict our considerations
to quasipoints $\frb \in \kQ(\mm \bar{\ten} \kL(\kh_{0}))$ that
contain $I_{\mm} \ten \pcx$. If $\frb$ is such a quasipoint, then
$I_{\mm} \ten \el \notin \frb$ for $\gl < 1$, since $\el \we \pcx = 0$.
Hence 
\[
    f_{I_{\mm} \ten A}(\frb) = 1,
\]
and this implies that the open set $\kQ_{I_{\mm} \ten \pcx}(\mm
\bar{\ten} \kL(\kh_{0}))$ is contained in $\urb{f}_{I_{\mm} \ten
A}(1)$. \ \ $\Box$ \\
~\\
We proceed with the development of the general theory.

\begin{proposition}\label{prop: 19}
    $f_{A} : \dr \to \RR$ is upper semicontinuous.
\end{proposition}
\emph{Proof:} We have to show that the following property holds:
\[
    \all \ \kj_{0} \in \dr \ \all \ \eps > 0 \ \ex \ P \in \kj_{0} \
    \all \ \kj \in \dpr : \ f_{A}(\kj) < f_{A}(\kj_{0}) + \eps. 
\]
Indeed $P := \ea_{f_{A}(\kj_{0}) + \frac{\eps}{2}} \in \kj_{0}$ and
therefore
\[
    f_{A}(\kj) ≤ f_{A}(\kj_{0}) + \frac{\eps}{2} < f_{A}(\kj_{0}) + \eps 
\]
for all $\kj \in \dpr$. \ \ $\Box$

\begin{remark}\label{rem: 20}
    Observable functions $f_{A} : \dr \to \RR$ are not continuous in
    general.
\end{remark}
\emph{Proof:} The observable function $f_{P} : \dr \to \RR$ of a
projection $P \in \rr$ is given by 
\[
    f_{P} = 1 - \chi_{\dpr}.
\]
$f_{P}$ is continuous if and only if $\dpr$ is open (which is true by 
definition) and closed. Now
\begin{eqnarray*}
    \kj \in \overline{\dpr} & \llra & \all \ Q \in \kj : \kD_{Q}(\rr) \cap 
    \dpr \neq \lm  \\
     & \llra & \all \ Q \in \kj : \kD_{P \we Q}(\rr) \neq \lm  \\
     & \llra & \all \ Q \in \kj : P \we Q \neq 0
\end{eqnarray*}
and therefore $\overline{\dpr} = \dpr$ if and only if
\[
    \all \ \kj \in \dr : \ (( \all \ Q \in \kj : \ P \we Q \neq 0) \
    \lra \ P \in \kj).
\]
This leads to the following example. Let $P, P_{1} \in \pr$ such that 
$0 \neq P < P_{1}$. Then $Q \we P = P$ for all $Q \in H_{P_{1}}$ but
$P \notin H_{P_{1}}$. Hence $H_{P_{1}} \in \overline{\dpr} \smm \dpr$.
\ \ $\Box$

\begin{proposition}\label{prop: 21}
    For any function $f : \dr \to \RR$, the following two properties are
    equivalent:
    \begin{enumerate}
	\item  [(i)] $f$ is upper semicontinuous and decreasing (i.e.
	$\kj_{1} \tm \kj_{2} \ \lra \ f(\kj_{2}) ≤ f(\kj_{1})$).  
    
	\item  [(ii)] $\all \ \kj \in \dr  : \ f(\kj) = \inf \{
	f(H_{P}) | \ P \in \kj \}$.
    \end{enumerate}
 \end{proposition}
\emph{Proof:} Assume that $(i)$ holds. Let $\kj_{0} \in \dr$ and $P
\in \kj_{0}$. Then $f(\kj_{0}) ≤ f(H_{P})$ and therefore $f(\kj_{0}) ≤
\inf \{ f(H_{P}) | \ P \in \kj_{0} \}$. Let $\eps > 0$. Then by the
upper semicontinuity of $f$
\[
    \ex \ P_{0} \in \kj_{0} \ \all \ \kj \in \kD_{P_{0}}(\rr) : \
    f(\kj) < f(\kj_{0}) + \eps,
\]
in particular
\[
    f(H_{P_{0}}) < f(\kj_{0}) + \eps.
\]
Hence $f(\kj_{0}) = \inf \{ f(H_{P}) | \ P \in \kj_{0} \}$. \\
Conversely assume that $(ii)$ holds. Let $\kj_{1}, \kj_{2} \in \dr, \ 
\ \kj_{1} \tm \kj_{2}$. Then
\[
    f(\kj_{2}) = \inf \{ f(H_{P}) | \ P \in \kj_{2} \} ≤ \inf \{ f(H_{P}) | \ P \in
    \kj_{1} \} = f(\kj_{1}),
\]
i.e. $f$ is decreasing. Let $\kj_{0} \in \dr$ and $\eps > 0$. There is
some $P \in \kj_{0}$ such that $f(H_{P}) < f(\kj_{0}) + \eps$. If $\kj
\in \dpr$ then $H_{P} \tm \kj$ and therefore
\[
    f(\kj) ≤ f(H_{P}) < f(\kj_{0}) + \eps.
\]
$\Box$

\begin{corollary}\label{cor: 22}
    $f_{A}(\dr) = sp(A)$ for all $A \in \hr$.
\end{corollary}
\emph{Proof:} We already know that $f_{A}(\qr) = sp(A)$, so it
suffices to prove that $f_{A}(\dr)$ is contained in $sp(A)$. But this 
follows from propositions \ref{prop: 16}, \ref{prop: 19}, \ref{prop:
21} and the closedness of $sp(A)$:
\[
    \all \ \kj \in \dr : \ f_{A} (\kj) = \inf_{P \in \kj}\sup_{\frb
    \in \qpr}f_{A}(\frb) \in sp(A). \ \ \Box
\]

\begin{definition}\label{def: 23}
    A function $f : \dr \to \RR$ is called an {\bf abstract observable
    function} if it is \emph{upper semicontinuous} and satisfies the
    \emph{intersection condition}
    \[
	f(\bigcap_{j \in J}\kj_{j}) = \sup_{j \in J}f(\kj_{j}) 
    \]
    for all families $(\kj_{j})_{j \in J}$ in $\dr$.
\end{definition}

The intersection condition implies that an abstract observable
function is decreasing. Hence by \ref{prop: 21} the definition of abstract
observable functions can be reformulated as follows:

\begin{remark}\label{rem: 23a}
    $f : \dr \to \RR$ is an observable function if and only if the
    following two properties hold for $f$:
    \begin{enumerate}
	\item  [(i)]  $\all \ \kj \in \dr  : \ f(\kj) = \inf \{ f(H_{P}) | \ P \in \kj
	\}$,
    
	\item  [(ii)] $f(\bigcap_{j \in J}\kj_{j}) = \sup_{j \in J}f(\kj_{j})$
	for all families $(\kj_{j})_{j \in J}$ in $\dr$.
    \end{enumerate}
\end{remark}

A direct consequence of the intersection condition is the following

\begin{remark}\label{rem: 24}
    Let $\gl \in im f$. Then the inverse image $\overset{-1}{f}(\gl) \tm
    \dr$ has a minimal element $\kj_{\gl}$ which is simply given by
    \[
	\kj_{\gl} = \bigcap \{ \kj \in \dr | \ f(\kj) = \gl \}.
    \]   
\end{remark}

We will now show how one can recover the spectral family $\ea$ of $A
\in \hr$ from the observable function $f_{A}$. This gives us the
decisive hint for the proof that to each abstract observable function 
$f : \dr \to \RR$ there is a unique $A \in \hr$ with $f = f_{A}$.
    
\begin{lemma}\label{lem: 25}
Let $ f_{A} : \dr \to  \RR$ be an observable function and let $\ea$ 
be the spectral family corresponding to $A$. If $\gl \in im f$, then 
\[
     \kj_{\gl} = \{ P \in \pr \mid \exists \  \mu > \gl : P ≥
     \ea_{\mu} \}.  
\]
$\kj_{\gl} = H_{\eal}$ if and only if $\ea$ is constant on some
interval $[\gl, \gl + \delta ]$. Moreover
\[
    \eal = \inf \kj_{\gl}.
\]
\end{lemma}
\emph{Proof}: It is obvious that $\kI := \{ P \in \dr \mid 
\exists \  \mu > \gl : P ≥ \ea_{\mu} \}$ is a dual ideal and that 
$f_{A}(\kI) = \gl$. Let $\kj$ be any dual ideal in $\dr$  such that 
$f_{A}(\kj) = \gl$. Then $\gl = \inf \{ \mu \mid \ea_{\mu} \in \kj \}$. 
For $P \in \kI $ let $\mu > \gl$ such that $P ≥ \ea_{\mu}$. Then 
$\ea_{\mu} \in \kj$ and therefore $P \in \kj$. This shows $\kI \tm \kj$. 
Hence $\kj_{\gl} = \kI.$ Clearly $\kj_{\gl} = H_{\eal}$ if and 
only if $\ea$ is constant on some interval $[\gl, \gl + \delta ]$. The
last assertion is due to the continuity of $\ea$ from the right.
{\bf $\Box$} \\

Let $\gl_{0} \in im f_{A}$. Then 
\[
    f_{A}(H_{\ea_{\gl_{0}}}) = \gl_{0}
\]
if and only if there is no $\gd > 0$ such that $\ea$ is constant on
the interval $[\gl_{0} - \gd, \gl_{0}]$.

\begin{proposition}\label{prop: 17}
    Let $\gl_{0} \in im f_{A}$. Then $\gl_{0} = f_{A}(H)$ for some
    principal dual ideal $H \in \dr$ if and only if there is no $\gd > 0$
    such that $\ea$ is constant on the interval $[\gl_{0} - \gd, \gl_{0}]$. 
\end{proposition}
\emph{Proof:} Assume that $\ea$ is constant on some interval
$[\gl_{1}, \gl_{0}]$ with $\gl_{1} < \gl_{0}$ but that there is a $P
\in \por$ such that $f_{A}(H_{P}) = \gl_{0}$. Then $\kj_{\gl_{0}} \tm 
H_{P}$ and therefore, by lemma \ref{lem: 25}, $\ea_{\gl_{0}} ≥ P$, i.e.
$H_{\ea_{\gl_{0}}} \tm H_{P}$. This implies $f_{A}(H_{\ea_{\gl_{0}}}) ≥
f_{A}(H_{P}) = \gl_{0}$, contradicting $f_{A}(H_{\ea_{\gl_{0}}}) ≤
\gl_{1} < \gl_{0}$. \ \ $\Box$ \\

\begin{corollary}\label{cor: 17a}
    If $\gl \in f_{A}(\dprr)$ then
    \[
	\eal = \Ve \{ P \in \por \ | \ f_{A}(H_{P}) = \gl \}. 
    \]    
\end{corollary}
\emph{Proof:} Note that, by proposition \ref{prop: 17}, the case $\eal
= 0$ cannot occur. If $f_{A}(H_{P}) = \gl$ then $\kj_{\gl} \tm H_{P}$ and
therefore $\eal ≥ P$. Thus $\gl = f_{A}(H_{\eal}) ≥ f_{A}(H_{P}) = \gl$. \
\ $\Box$ \\

\begin{corollary}\label{cor: 18}
    An observable function $f_{A} : \dr \to \RR$ is uniquely
    determined by its restriction to $\qr$.
\end{corollary}
\emph{Proof:} By proposition \ref{prop: 21} $f_{A}$ is determined by its
values on principle dual ideals $H_{P} \ (P \in \pr)$. $H_{P} = \bigcap
\qpr$ by lemma \ref{lem: 15} and therefore $f_{A}(H_{P}) = 
\sup \{ f_{A}(\frb) | \ \frb \in \qpr \}$ by proposition \ref{prop: 16}.
\ \ $\Box$ \\

\begin{remark}\label{rem: 25a}
    A selfadjoint operator $A \in \rr$ is uniquely determined by its
    observable function $f_{A}$.
\end{remark}
\emph{Proof:} If $\gl \in im f_{A}$ then $\eal = \inf \kj_{\gl}$ and
$\kj_{\gl}$ is the minimal element of $\overset{-1}{f_{A}}(\gl)$.
Hence the uniqueness of $A$ follows from the uniqueness of the
spectral resolution. \ \ $\Box$ \\

\begin{theorem}\label{theo: 26}
    Let $f : \dr \to \RR$ be an abstract observable function. Then
    there is a unique $A \in \hr$ such that $f = f_{A}$.
\end{theorem}
The {\bf proof} will proceed in three steps. In the first step we
construct from the abstract observable function $f$ an increasing family
$(\el)_{\gl \in im f}$ in $\pr$ and show in a second step that this
family can be extended to a spectral family in $\rr$. Finally, in the 
third step, we show that the selfadjoint operator $A \in
\rr$ corresponding to that spectral family has observable function
$f_{A} = f$ and that $A$ is uniquely determined by $f$.

\paragraph{Step 1}
    
Let $\gl \in im f$ and let $\jl \in \dr$ be the smallest dual ideal
such that $f(\jl) = \gl$. In view of lemma \ref{lem: 25} we have no
choice than to define
\[
    \el := \inf \jl.
\]
\begin{lemma}\label{lem: 27}
    The family $(\el)_{\gl \in im f}$ is increasing.
\end{lemma}
\emph{Proof:} Let $\gl, \mu \in im f, \ \gl < \mu$. Then
\begin{eqnarray*}
    f(\jm) & = & \mu  \\
     & = & max(\gl, \mu)   \\
     & = & max(f(\jl), f(\jm))  \\
     & = & f(\jl \cap \jm).
 \end{eqnarray*}
Hence, by the minimality of $\jm$,
\[
    \jm \tm \jl \cap \jm \tm \jl
\]
and therefore $\el ≤ \emm$. \ \ $\Box$

\begin{lemma}\label{lem: 28}
    $f$ is monotonely continuous, i.e. if $(\kj_{j})_{j \in J}$ is an 
    increasing net in $\dr$ then 
    \[
	f(\bigcup_{j \in J}\kj_{j}) = \lim_{j}f(\kj_{j}).
    \]
\end{lemma}
\emph{Proof:} Obviously $\kj := \bigcup_{j \in J}\kj_{j} \in \dr$. 
As $f$ is decreasing, $f(\kj) ≤ f(\kj_{j})$ for all $j \in J$ and
$(f(\kj_{j})_{j \in J}$ is a decreasing net of real numbers. Hence
\[
    f(\kj) ≤ \lim_{j}f(\kj_{j}).
\]
Let $\eps > 0$. Because $f$ is upper semicontinuous, there is $P \in
\kj$ such that $f(\kI) < f(\kj) + \eps$ for all $\kI \in \dpr$. Now $P
\in \kj_{k}$ for some $k \in J$ and therefore
\[
     \lim_{j}f(\kj_{j}) ≤ f(\kj_{k}) < f(\kj) + \eps,
\]
which shows that also $\lim_{j}f(\kj_{j}) ≤ f(\kj)$ holds. \ \
$\Box$

\begin{corollary}\label{cor: 29}
    The image of an abstract observable function is compact.
\end{corollary}
\emph{Proof:} Because $\{I\} \tm \kj$ for all $\kj \in \dr$ we have $f
≤ f(\{I\})$ on $\dr$.\\
If $\gl, \mu \in im f$ and $\gl < \mu$ then $\jm \tm \jl$, hence
$\bigcup_{\gl \in im f}\jl$ is a dual ideal and therefore contained in
a maximal dual ideal $\frb \in \dr$. This shows $f(\frb) ≤ f$ on $\dr$
and consequently $im f$ is bounded. Let $\gl \in \overline{im f}$.
Then there is an increasing sequence $(\mu_{n})_{n \in \NN}$ in $im f$
converging to $\gl$ or there is a decreasing sequence $(\mu_{n})_{n \in
\NN}$ in $im f$ converging to $\gl$. In the first case we have
$\kj_{\mu_{n + 1}} \tm \kj_{\mu_{n}}$ for all $n \in \NN$ and
therefore for $\kj := \bigcap_{n}\kj_{\mu_{n}} \in \dr$
\[
    f(\kj) = \sup_{n}f(\kj_{\mu_{n}}) = \sup_{n}\mu_{n} = \gl.
\] 
In the second case we have $\kj_{\mu_{n}} \tm \kj_{\mu_{n + 1}}$
for all $n \in \NN$ and therefore $\kj := \bigcup_{n}\kj_{\mu_{n}} \in \dr$.
Hence
\[
    f(\kj) = \lim_{n}f(\kj_{\mu_{n}}) = \lim_{n}\mu_{n} = \gl.
\]
Therefore $\gl \in im f$ in both cases, i.e. $im f$ is also closed.
\ \ $\Box$

\paragraph{Step 2}

We will now extend $(\el)_{\gl \in im f}$ to a spectral family
$\ef := (\el)_{\gl \in \RR}$. In defining $\ef$ we have of course in
mind that the spectrum of the selfadjoint operator $A$ corresponding
to $\ef$ should coincide with $im f$. This forces us to define $\el$
for $\gl \notin im f$ in the following way. For $\gl \notin im f$ let
\[
    S_{\gl} := \{ \mu \in im f | \ \mu < \gl \}.
\]  
Then we define
\[
    \el :=
    \begin{cases}
	0     &  \text{if} \ \  S_{\gl} = \emptyset  \\
	E_{sup \  S_{\gl}}    &   \text{otherwise}.
    \end{cases}    
\]
Note that $f(\{ I \}) = max \ im f$ and that $\kj_{f(\{ I \})} = \{ I
\}$.
\begin{lemma}\label{lem: 30}
    $\ef$ is a spectral family.
\end{lemma}
\emph{Proof:} The only remaining point to prove is that $\ef$ is
continuous from the right, i.e. that $\el = \We_{\mu > \gl}\emm$ for
all $\gl \in \RR$. This is obvious if $\gl \notin im f$ or if there is
some $\gd > 0$ such that $]\gl, \gl + \gd[ \cap im f =\emptyset$.
Therefore we are left with the case that there is a strictly decreasing
sequence $(\mu_{n})_{n \NN}$ in $im f$ converging to $\gl$. For all $n
\in \NN$ we have $f(\kj_{\mu_{n}}) > f(\kj_{\gl})$ and therefore
$\kj_{\mu_{n}} \tm \kj_{\gl}$. Hence $\bigcup_{n}\kj_{\mu_{n}} \tm
\kj_{\gl}$ and 
\[
     f(\bigcup_{n}\kj_{\mu_{n}}) = \lim_{n}f(\kj_{\mu_{n}}) = \gl
\]
implies $\bigcup_{n}\kj_{\mu_{n}} = \kj_{\gl}$ by the minimality of
$\kj_{\gl}$. If $P \in \kj_{\gl}$ then $P \in \kj_{\mu_{n}}$ for some 
$n$ and therefore $E_{\mu_{n}} ≤ P$. This shows $\We_{\mu > \gl}\emm ≤
P$. As $P \in \kj_{\gl}$ is arbitrary we can conclude that $\We_{\mu >
\gl}\emm ≤ \el$. The reverse inequality is obvious. \ \ $\Box$ \\ 

\paragraph{Step 3}

 
Let $A \in \rr$ be the selfadjoint operator corresponding to the
spectral family $\ef$. It is obvious from the definition of $\ef$ that
\[
    sp(A) \tm im f.
\]  
The next result shows that the spectrum of $A$ is equal to the image
of $f$.
\begin{lemma}\label{lem: 31}
    Let $\el$ be constant on the nonempty interval $]\gl_{0},
    \gl_{1}[$.
    Then 
    \[
	im f \cap ]\gl_{0}, \gl_{1}[ = \emptyset. 
    \]
\end{lemma}
\emph{Proof:} Because of the right-continuity of $\ef$ we can assume that
$\gl_{0}$ belongs to $im f$. We show first that $im f \cap ]\gl_{0},
\gl_{1}[$ consists of at most one element.\\
Assume that $\gl, \mu \in im f, \ \ \gl_{0} < \gl < \mu < \gl_{1}.$
Because $f$ is upper semicontinuous we can find, given $\eps \in ]0,
\gl - \gl_{0}[$, a projection $P \in \kj_{\gl_{0}}$ such that
\[
    \all \ \kj \in \dpr : \ f(\kj) < \gl_{0} + \eps < \gl,
\]
in particular
\[
    \gl_{0} = f(\kj_{\gl_{0}}) ≤ f(H_{P}) < \gl,
\]
hence $\kj_{\gl} \tm H_{P} \tm \kj_{\gl_{0}}$. This implies
\[
    P = \inf \ H_{P} ≤ \inf \ \kj_{\gl} = \el = E_{\gl_{0}},
\]
hence $E_{\gl_{0}} = P \in \kj_{\gl_{0}}$ and therefore
\[
    \kj_{\gl_{0}} = H_{E_{\gl_{0}}}.
\]
By the same argument, applied to $\gl$ and $\mu$, we see that
\[
    \kj_{\gl} = H_{\el} = H_{E_{\gl_{0}}} = \kj_{\gl_{0}}
\]
and therefore
\[
    \gl_{0} = f(\kj_{\gl_{0}}) = f(\kj_{\gl}) = \gl,
\]
a contradiction. \\
We now show that $\gl_{0} \in im f$ and $]\gl_{0}, \gl[ \cap im f =
\emptyset$ imply $\kj_{\gl_{0}} = H_{E_{\gl_{0}}}$.\\
Choose $\eps > 0$ sufficiently small and choose $P \in \kj_{\gl_{0}}$ 
such that
\[
    \all \ \kj \in \dpr : \ f(\kj) < f(\kj_{\gl_{0}}) + \eps < \gl.
\]
Then $]\gl_{0}, \gl[ \cap im f = \emptyset$ implies
\[
    \all \ \kj \in \dpr : \ f(\kj) ≤ \gl_{0},
\]
in particular 
\[
    \gl_{0} = f(\kj_{\gl_{0}}) ≤ f(H_{P}) ≤ \gl_{0}.
\]
This shows $ \kj_{\gl_{0}} \tm H_{P}$ and therefore 
\[
    P = \inf \ H_{P} ≤ \inf \ \kj_{\gl_{0}} = E_{\gl_{0}}.
\]
Because of $P \in \kj_{\gl_{0}}$ we therefore have $E_{\gl_{0}} \in
\kj_{\gl_{0}}$, i.e. $\kj_{\gl_{0}} = H_{E_{\gl_{0}}}$.\\
Finally assume that $]\gl_{0}, \gl_{1}[ \cap im f \neq \emptyset$ and 
let $\gl$ be the unique element of this intersection. Then by the
foregoing we obtain $\kj_{\gl_{0}} = H_{E_{\gl_{0}}} = H_{E_{\gl}} =
\kj_{\gl}$, i.e. $\gl_{0} = \gl$, a contradiction. \ \ $\Box$ \\

\begin{corollary}\label{cor: 32}
    Let $f : \dr \to \RR$ be an abstract observable function and let$A
    \in \rr$ be the selfadjoint operator corresponding to the spectral
    family $\ef$ defined by $f$. Then $sp(A) = im f$.
\end{corollary}
\emph{Proof:} Note that $\gl \notin sp(A)$ if and only if $\ef$ is
constant on some neighborhood of $\gl$. The definition of $\ef$ shows 
that $\RR \smm im f \tm \RR \smm sp(A)$, i.e. $sp(A) \tm im f$, and
the foregoing lemma shows that $\RR \smm sp(A) \tm \RR \smm im f$, i.e.
$ im f \tm sp(A)$. \ \ $\Box$ \\

\begin{lemma}\label{lem: 33}
    Let $f : \dr \to \RR$ be an abstract observable function, $A$ the 
    selfadjoint operator defined by $f$ and $f_{A} : \dr \to \RR$ the 
    observable function corresponding to $A$. Then $f_{A} = f$.
\end{lemma}
\emph{Proof:} Recall that $f_{A}(\kj) = \inf \{ \gl | \ \el \in \kj \}$
for all $\kj \in \dr$. Due to $sp(A) = im f_{A} = im f$ this can be
written as $f_{A}(\kj) = \inf \{ \gl \in sp(A) | \ \el \in \kj \}$. 
If $\el \in \kj$ (with $\gl \in sp(A)$) then $\kj_{\gl} \tm H_{\el}
\tm \kj$ and therefore $f(\kj) ≤ f(\kj_{\gl}) = \gl$. This implies
\[
    f ≤ f_{A}.
\]
For the proof of the reverse inequality we distinguish two cases. \\

{\bf (i)}  Let $\kj \in \dr$ and let $\gl = f(\kj_{\gl}) = f(\kj)$ be 
isolated from the right, i.e. $im f \cap ]\gl, \mu[ = \emptyset$ for
some $\mu > \gl$. Then, by the proof of the foregoing lemma, $\el \in 
\kj_{\gl} \tm \kj$ and therefore 
\[
    f_{A}(\kj) ≤ \gl = f(\kj).
\]   

{\bf (ii)}  Let $\gl = f(\kj_{\gl})$ be \emph{not} isolated from the
right and let $(\mu_{n})_{n \in \NN}$ be a strictly decreasing
sequence in $im f$ that converges to $\gl$. If $\el = \inf \kj_{\gl}
\in \kj_{\gl}$ then $f(\kj_{\gl}) = f_{A}(\kj_{\gl})$. Let $\el \notin
\kj_{\gl}$. Let $n \in \NN$. Because $f$ is upper semicontinuous
there is $P \in \kj_{\mu_{n + 1}}$ such that
\[
    \all \ \kj \in \dpr : \ f(\kj) < \mu_{n}.
\]  
In particular
\[
    \mu_{n + 1} ≤ f(H_{P}) < \mu_{n} = f(\kj_{\mu_{n}}).
\]
Now $f(H_{P} \cap \kj_{\mu_{n}}) = max(f(H_{P}), f(\kj_{\mu_{n}})) =
f(\kj_{\mu_{n}})$ and therefore $\kj_{\mu_{n}} \tm H_{P}$ by the
minimality of $\kj_{\mu_{n}}$. So we
obtain
\[
    E_{\mu_{n + 1}} ≤ P ≤ E_{\mu_{n}},
\]
hence
\[
E_{\mu_{n}} \in \kj_{\mu_{n + 1}} \tm \kj_{\gl}.
\]
This shows $E_{\gl + \eps} \in \kj_{\gl}$ for all $\eps > 0$ and
therefore $f_{A}(\kj_{\gl}) ≤ \gl$. This proves $f_{A} = f$. \ \
$\Box$ \\
~\\
The uniqueness of $A$ is obvious by remark \ref{rem: 25a}: if $A, B
\in \hr$ such that $f_{A} = f = f_{B}$ then $A = B$. \\
\emph{This completes the proof of theorem \ref{theo: 26}.} \\
~\\
The theorem confirms that there is no difference between ``abstract'' 
and ``concrete'' observable functions and therefore we will speak
generally of observable functions. \\
~\\
Let $\por$ denote the set of nonzero projections in $\rr$.
We will now show that observable functions can be characterized as
functions $ \por \to \RR$ that satisfy a ``continuous join condition''.
Note that for an arbitrary family $(P_{k})_{k \in \KK}$ in $\por$ we
have
\[
    \bigcap_{k \in \KK}H_{P_{k}} = H_{\Ve_{k \in \KK}P_{k}}. 
\]
If $f : \dr \to \RR$ is an observable function then the intersection
property implies
\[
     f(H_{\Ve_{k \in \KK}P_{k}}) = \sup_{k \in \KK}f(H_{P_{k}}).
\]
This leads to the following

\begin{definition}\label{def: 34}
    A bounded function $r : \por \to \RR$ is called completely increasing if
    \begin{equation}
	r(\Ve_{k \in \KK}P_{k}) = \sup_{k \in \KK}r(P_{k})
	\label{eq:10}
    \end{equation}
    for every family $(P_{k})_{k \in \KK}$ in $\por$.
\end{definition}
Note that it is sufficient to assume in the foregoing definition that 
$r$ is bounded from below because $r(I)$ is an upper bound, in fact
the maximum, for an arbitrary increasing function $r : \por \to \RR$. 
\\
Because of the natural bijection $P \tto H_{P}$ between $\por$ and the 
set $\dprr$ of principle dual ideals of $\pr$ each observable function
$f : \dr \to \RR$ induces by restriction a completely increasing function
$r_{f}$: 
\[
     \all \ P \in \por : \ r_{f}(P) := f(H_{P}).
\]
Conversely we will now show that each completely increasing function
on $\por$ induces an observable function so that we get a one to one
correspondence between observable functions and completely increasing 
functions. This will enable us to complete the proof of theorem
\ref{theo: 11}.\\

\begin{definition}\label{def: 35}
    Let $r : \por \to \RR$ be a completely increasing function. Then we
    define a function $f_{r} : \dr \to \RR$ by
    \[
	 \all \ \kj \in \dr : \ f_{r}(\kj) := \inf_{P \in \kj}r(P).
    \]
\end{definition}

\begin{remark}\label{35a}
    It is this definition where we need that $r : \por \to \RR$ is
    bounded from below. The following example shows that there are
    functions $r : \por \to \RR$ that satisfy the condition
    \ref{eq:10}, but are not bounded from below. \\
    ~\\
    Let $E = (\el)_{\gl \irr}$ be a spectral family that is bounded
    from above but not from below. Then $\el \ne 0$ for all $\gl \irr$.
    Let $M := \min \{ \gl \mid \el = I \}$. If $P \in \por$, then $\{ 
    \gl ≤ M \mid P ≤ \el \}$ is a bounded set, and we can define a function
    $r : \por \to \RR$ by
    \[
	r(P) := \inf \{ \gl \mid \el ≥ P \}. 
    \]
    It is easy to see that the proof of the intersection property for 
    observable functions also works in this case, so that we get
    \[
	r(\Ve_{k \in \KK}P_{k}) = \sup_{k \in \KK}r(P_{k})
    \]	    
    for every family $(P_{k})_{k \in \KK}$ in $\por$. But $r(\el) ≤
    \gl$ for all $\gl \irr$, so $r$ is not bounded from below.    
\end{remark}
It is obvious that
\[
    \all \ P \in \por : \ f_{r}(H_{P}) = r(P)
\]
holds.

\begin{proposition}\label{prop: 36}
    The function $f_{r} : \dr \to \RR$ induced by the completely increasing
    function $r : \por \to \RR$ is an observable function.
\end{proposition}
\emph{Proof:} In view of proposition \ref{prop: 21} we have to show
that $f_{r}$ satisfies
\[
    f_{r}(\bigcap_{k \in \KK}\kj_{k}) = \sup_{k \in \KK}f_{r}(\kj_{k})
\]
for all families $(\kj_{k})_{k \in \KK}$ in $\dr$. Since $f_{r}$ is
decreasing we have
\[
    f_{r}(\bigcap_{k \in \KK}\kj_{k}) ≥ \sup_{k \in \KK}f_{r}(\kj_{k}).
\]
Let $\eps > 0$ and choose $P_{k} \in \kj_{k} \ \ (k \in \KK)$ such
that $r(P_{k}) < f_{r}(\kj_{k}) + \eps$. Now $\bigcap_{k}H_{P_{k}}
\tm \bigcap_{k}\kj_{k}$, $f_{r}$ is decreasing and $r$ is completely
increasing, hence
\[
    f_{r}(\bigcap_{k}\kj_{k}) ≤ f_{r}(\bigcap_{k}H_{P_{k}}) =
    r(\Ve_{k}P_{k}) = \sup_{k}r(P_{k}) ≤ \sup_{k}f_{r}(\kj_{k}) + \eps
\]
and therefore 
\[
     f_{r}(\bigcap_{k \in \KK}\kj_{k}) ≤ \sup_{k \in \KK}f_{r}(\kj_{k}).
     \ \ \Box
\]
~\\
~\\
    We have formulated theorem \ref{theo: 26} and the
    characterization of observable functions by completely increasing 
    functions in the category of von Neumann algebras. A simple
    inspection of the proofs shows that we have used the fact that the
    projection lattice $\pr$ of a von Neumann algebra $\rr$ is a complete
    orthomodular lattice. Therefore we can translate theorem
    \ref{theo: 26} to the category of complete orthomodular lattices
    in the following way:
    
    \begin{theorem}\label{theo: 26a}
	Let $\LL$ be a complete orthomodular lattice and let $f : \dl \to \RR$
	be an abstract observable function. Then there is a unique
	spectral family $E$ in $\LL$ such that $f = f_{E}$.
    \end{theorem}
    The function $f_{E} : \dl \to \RR$ is defined quite naturally as
    \[
	\all \ \kj \in \dl : \ f_{E}(\kj) := \inf \{ \gl \irr \ | \
	\el \in \kj \}.
    \]
~\\ 
We will present now a further characterization of observable
functions. For a function $f : \por \to \RR$ let
\[
    \kf_{\gl} := \overset{-1}{f}(]-\∞, \gl]) \cup \{0\}.
\]
Note that $f$ is lower semicontinuous with respect to the topology of 
strong convergence if and only if $\kf_{\gl}$ is strongly closed in
$\pr$ for all $\gl \in \RR$.

\begin{proposition}\label{prop: 36a}
    Let $f : \por \to \RR$ be a function. Then the following
    properties are equivalent:
    \begin{description}
        \item[(i)]  $f$ is completely increasing.
    
        \item[(ii)]  $f$ is strongly lower semicontinuous and $f(P
	\vee Q) = \max (f(P), f(Q))$ for all $P, Q \in \por$ .
    
        \item[(iii)]  For all $\gl \in \RR$ the set $\kf_{\gl}$ is a strongly
	closed ideal in $\pr$.
    \end{description}
\end{proposition}
\emph{Proof:} Let $f$ be completely increasing and let $(P_{a})_{a
\in \AAA}$ be a net in $\mbox{$\overset{-1}{f}(]-\∞, \gl])$}$ that converges
strongly to $P \in \por$. Because of $P_{a} ≤ \Ve_{b \in \AAA}P_{b}$
for all $a \in \AAA$ we have also $P ≤ \Ve_{b \in \AAA}P_{b}$ and
therefore $f(P) ≤ f(\Ve_{b \in \AAA}P_{b}) = \sup \{f(P_{a}) \ | \ a
\in \AAA \} ≤ \gl$.\\
Now assume that (ii) holds. We have to show that
$\kf_{\gl}$ is an ideal in $\pr$ for all $\gl \irr$. If $Q \in
\kf_{\gl}$ and $P ≤ Q$ then $P \in \kf_{\gl}$ because $f$ is increasing.
If $P$ and $Q$ are two nonzero elements of $\kf_{\gl}$. Then $P \vee Q
\in \kf_{\gl}$ because of $f(P \vee Q) = \max (f(P), f(Q))$. \\
Finally we show that (iii) implies (i). Let $P, Q \in \por$ and $P ≤
Q$. From $Q \in \kf_{f(Q)}$ we conclude $P \in \kf_{f(Q)}$, i.e. $f(P)
≤ f(Q)$. Hence $f$ is increasing and therefore $f(P \vee Q) ≥ \max
(f(P), f(Q))$ for all $P, Q \in \por$. Now $P, Q \in \kf_{\max (f(P),
f(Q))}$ and therefore $P \vee Q \in \kf_{\max (f(P), f(Q))}$ because
$\kf_{\max (f(P), f(Q))}$ is an ideal. This shows $f(P \vee Q) ≤ \max (f(P),
f(Q))$. Now let $(P_{a})_{a \in \AAA}$be an arbitrary family in $\por$
and let $Q := \Ve_{b \in \AAA}P_{b}$. Then $Q$ is the strong limit of
the increasing net $(Q_{F})_{F \in Fin(\AAA)}$ where $Q_{F} := \Ve_{a 
\in F}P_{a}$ and $Fin(\AAA)$ denotes the set of all nonempty finite
subsets of $\AAA$. From $P_{a} \in \kf_{\sup_{b \in \AAA}f(P_{b})}$ for 
all $a \in \AAA$, we obtain $Q_{F} \in \kf_{\sup_{b \in
\AAA}f(P_{b})}$ for all $F \in Fin(\AAA)$ and therefore, as $\kf_{\sup_{b
\in \AAA}f(P_{b})}$ is strongly closed, $Q \in \kf_{\sup_{b \in
\AAA}f(P_{b})}$. Hence $f(\Ve_{a \in \AAA}P_{a}) = \sup_{a \in
\AAA}f(P_{a})$. \ \ $\Box$ \\
~\\
If $\rr$ is a von Neumann algebra, $r : \por \to \RR$ a completely 
increasing function, $\gl \in im \ r$ and
\[
	E_{\gl} := \Ve \{ P \in \por \ | \ r(P) = \gl \}
\] 
then also $r(E_{\gl}) = \gl$. Hence $E_{\gl}$ is the largest
element in the inverse image $\overset{-1}{r}(\gl)$. It is easy to
see that
\begin{enumerate}
    \item  [(i)] $\gl, \mu \in im \ r$ and $\gl ≤ \mu$ imply $E_{\gl}
	≤ E_{\mu}$,
    
    \item  [(ii)] and if $(\mu_{n})_{n \inn}$ is a decreasing
	sequence in $im \ r$ converging to $\gl \in im \ r$ then
	$E_{\gl} = \We_{n \inn}E_{\mu_{n}}$       
\end{enumerate}
hold. Let $f : \dr \to \RR$ the observable function induced by $r$,
$\ea = (\eal)_{\lir}$ the corresponding spectral family and $A \in \hr$
the selfadjoint operator defined by $\ea$.  

\begin{remark}\label{36b}
    The range of $r$ is dense in $sp(A)$: $sp(A) = \overline{r(\por)}$.
\end{remark}
\emph{Proof:} This follows from $f(\dr) = sp(A)$ and $f(\kj) = \inf_{P
\in \kj}r(P)$ for all $\kj \in \dr$. \ \ $\Box$ \\
~\\
For $\gl \notin sp(A)$ we define $\el$ in the very same way as in step
2 of the proof of theorem \ref{theo: 26}. Then $\frpp_{r} := 
(\el)_{\gl \in r(\por) \cup (\RR \smm sp(A))}$ becomes a prespectral
family. From corollary \ref{cor: 17a} we know that 
\[
	\all \ \gl \in r(\por) : \ \el = \eal.
\]
Hence the foregoing remark and property $(ii)$ show that the
spectralization $E$ of $\frpp_{r}$ coincides with the spectral family 
$\ea$. So we have proved

\begin{proposition}\label{36c}
    Let $r : \por \to \RR$ be a completely increasing function and
    let $A \in \hr$ be the selfadjoint operator determined by $r$.
    Then the spectral family $\ea$ of $A$ is the unique extension of
    the family $(\el)_{\gl \in r(\por)}$, defined by
    \[
	 E_{\gl} := \Ve \{ P \in \por \ | \ r(P) = \gl \}.
    \]
\end{proposition}
~\\
~\\
The special case $\rr = \lh$ deserves a detailed study. Let $\ph$ the 
lattice of projections in $\lh$, $\poh$ the subset of nonzero
projections and $\puh$ the subset of projections of rank one. The
decisive feature of the special case $\lh$ is that every element of
$\poh$ is the join of a suitable family in $\puh$.  \\
If $r : \poh \to \RR$ is a completely increasing function then clearly
$r$ is uniquely determined by its restriction to $\puh$. Of course not
every function $s : \puh \to \RR$ is the restriction of a completely
increasing function on $\poh$: because the representation of $P \in
\poh$ as the join of a family in $\puh$ is far from being unique in
general, $s$ must satisfy some compatibility condition (and, as it
turns out, some continuity condition too). \\
The compatibility condition is easy to detect: let $P_{1}, P_{2}$ be
two different elements of $\puh$ and let $Q \in \puh$ be a
subprojection of $P_{1} \vee P_{2}$. Then necessarily
\[
    r(Q) ≤ r(P_{1} \vee P_{2}) = \max (r(P_{1}), r(P_{2})).
\]
Hence the restriction $s : \puh \to \RR$ of a completely increasing
function $r : \poh \to \RR$ must satisfy 
\[
    \all \ P, Q, R \in \puh : \ (P ≤ Q \vee R \ \ \lra \ \ s(P) ≤ \max
    (s(Q), s(R)).
\]

\begin{lemma}\label{lem: 36a}
    Let $r : \poh \to \RR$ be a completely increasing function. Then
    the restriction $s$ of $r$ to $\puh$ is lower semicontinuous with 
    respect to the topology of strong convergence on $\puh$.
\end{lemma}
\emph{Proof:} We have to show that for every $\gl_{0} \in \RR$
\[
    \overset{-1}{s}(]- \∞, \gl_{0}]) = \{ P \in \puh \  | \  s(P) ≤ \gl_{0} 
    \}
\]
is closed with respect to strong convergence. Let $f$ be the
observable function induced by $r$ and let $(\el)_{\gl \in \RR}$ be
the spectral family corresponding to $f$. If $P \in \poh$ then $r(P) =
f(H_{P}) ≤ \gl_{0}$ if and only if $E_{\gl_{0}} ≥ P$. Now consider a
net $(P_{k})_{k \in \KK}$ in $\overset{-1}{s}(]-\∞, \gl_{0}]$ that
converges strongly to $P \in \puh$. Then $E_{\gl_{0}}P_{k} = P_{k}$
for all $k \in \KK$ and as $E_{\gl_{0}}P_{k} \to E_{\gl_{0}}P$
strongly we conclude $E_{\gl_{0}}P =P$, i.e. $P \in \overset{-1}{s}(]-\∞,
\gl_{0}]$. \ \ $\Box$ \\

We say that a function $s : \puh \to \RR$ \emph{induces} a completely 
increasing function $r : \poh \to \RR$ if the restriction of $r$ to
$\puh$ is $s$.

\begin{theorem}\label{theo: 36b}
    A bounded function $s : \puh \to \RR$ induces a completely 
    increasing function $r : \poh \to \RR$ if and only if the following
    two conditions are satisfied:
    \begin{enumerate}
	\item  [(i)] $s$ is lower semicontinuous with respect to the
	topology of strong convergence on $\puh$,
    
	\item  [(ii)] $s(P) ≤ \max (s(Q), s(R))$ for all $P, Q, R \in 
	\puh$ such that $P ≤ Q \vee R$.
    \end{enumerate}
\end{theorem}
\emph{Proof:} A completely increasing function is bounded and
therefore its restriction to $\puh$ must be bounded too. We have already
seen that the conditions (i) and (ii) are necessary. \\
Conversely, assume that they are fullfilled for a bounded function 
$s : \puh \to \RR$. Let $Q \in \poh$ and let $(P_{k})_{k \in \KK}$ be 
a family in $\puh$ such that $Q = \Ve_{k \in \KK}P_{k}$. Then we are 
forced to define
\[
    r(Q) := \sup_{k \in \KK}s(P_{k}).
\] 
In order to show that this is well-defined we begin with the case that
$\KK$ is a finite set.\\
    Let $\KK$ be a finite non-empty set and let $Q = \Ve_{k \in \KK}P_{k}$
    with $P_{k} \in \puh$ for all $k \in \KK$. Then
    \[
	\sup \{s(P) \ | \ P \in \puh, \ P ≤ Q \} = \max_{k \in \KK}s(P_{k}).  
    \]
For the proof we use induction with respect to the size $n$
of $\KK$. For $n = 1$ there is nothing to prove. Let $n > 1$ and
assume that the claim is true for all subsets of $\KK$ of size $n - 1$.
Let $P_{k} = P_{\CC x_{k}} \ (k \in \KK), \  P = P_{\CC x}$ with $x,
x_{k} \in S^1(\kh)$. Obviously $x = \sum_{k = 1}^{n - 1}a_{k}x_{k} +
a_{n}x_{n}$ with suitable $a_{1}, \ldots, a_{n} \in \CC$ and therefore
$P ≤ P_{\CC \sum_{k = 1}^{n - 1}a_{k}x_{k}} \vee P_{n}$. By the
induction hypothesis we have
\[
    s(P) ≤ \max (s(P_{\CC \sum_{k = 1}^{n - 1}a_{k}x_{k}}), s(P_{n})) ≤
    \max_{k ≤ n}s(P_{k}).
\]
Hence $\sup_{P ≤ Q}s(P) ≤ \max_{k ≤ n}s(P_{k})$ and the opposite
inequality is trivial. \\

Now let $\KK$ be an arbitrary non-empty set, $P = P_{\CC x} ≤ Q$ and
let $Fin(\KK)$ be the set of all finite non-empty subsets of $\KK$.
Then $Q = \Ve_{F \in Fin(\KK)}Q_{F}$, where $Q_{F} := \Ve_{j \in
F}P_{j}$, $x$ is the limit of a net of unit vectors $x_{F} \in
lin_{\CC}\{x_{j} \ | \ j \in F \}$ and therefore the net $(P_{\CC
x_{F}})$ is strongly convergent to $P$. From the finite case we obtain
\[
    s(P_{\CC x_{F}}) ≤ \max_{j \in F}s(P_{j}) ≤ \gl_{0} := \sup_{k \in
    \KK}s(P_{k}).
\]        
Hence $P_{\CC x_{F}} \in \overset{-1}{s}(]-\∞, \gl_{0}]$ for all $F
\in Fin(\KK)$ and therefore $s(P) ≤ \gl_{0}$ by the lower
semicontinuity of $s$. This shows $\sup \{s(P) \ | \ P \in \puh, \ P ≤ Q \}
= \sup_{k \in \KK}s(P_{k})$. Thus $r$ is well-defined and obviously
completely increasing. \ \ $\Box$ \\  
~\\
~\\
Now we will finish the proof of theorem \ref{theo: 11}. Let $\rr$ be an abelian
von Neumann algebra and let $f : \qr \to \RR$ be a continuous function.
($f$ is necessarily bounded because $\rr$ is abelian and therefore,
due to Stone's theorem, $\qr$ is compact.)
Then, using corollary I.3.1 and theorem I.3.1, we have for an
arbitrary family $(P_{k})_{k \in \KK}$ in $\por$
\begin{eqnarray*}
    \sup \{ f(\frb) | \frb \in \kQ_{\Ve_{k \in \KK}P_{k}}(\rr) \} & = & \sup
    \{ f(\frb) | \frb \in \overline{\bigcup_{k \in \KK}\kQ_{P_{k}}(\rr)}
   \} \\
     & = & \sup \{ f(\frb) | \frb \in \bigcup_{k \in \KK}\kQ_{P_{k}}(\rr)
   \}  \\
     & = & \sup_{k} \sup \{ f(\frb) | \frb \in \kQ_{P_{k}}(\rr) \}.
\end{eqnarray*}
Because of 
\[
    H_{P} = \bigcap \qpr
\]
for all $P \in \por$ it is natural to define
\[
    r(P) := \sup \{ f(\frb) | \frb \in \qpr \}.
\]
Then we obtain from the foregoing computation
\[
    r(\Ve_{k \in \KK}P_{k}) = \sup_{k \in \KK}r(P_{k}),
\]
i.e. $r : \pr \to \RR$ is a completely increasing function.
Let $f_{r} : \dr \to \RR$ be the corresponding observable function.
The following lemma completes the proof of theorem \ref{theo: 11}.

\begin{lemma}\label{lem: 39}
    $f$ coincides with the restriction of $f_{r}$ to $\qr$.
\end{lemma}
\emph{Proof:} We have to show that
\[
    \all \ \frb \in \qr : \ f(\frb) = \inf_{P \in \frb}r(P) 
\]
holds. \\
From the definition of $r$ we see that $f(\frb) ≤ m := \inf_{P \in
\frb}r(P)$. Let $\eps > 0$. Because $f$ is continuous there is $P_{0} 
\in \frb$ such that $f(\frc) < f(\frb) + \eps$ on $\kQ_{P_{0}}(\rr)$. 
Hence 
\[
    m ≤ r(P_{0}) = \sup f(\kQ_{P_{0}}(\rr)) ≤ f(\frb) + \eps.
\]   
This shows $m ≤ f(\frb)$. \ \ $\Box$
\vspace{5mm}

In the next subsection we will show that for abelian von Neumann
algebras the bijection $A \tto f_{A}$ from $\rr$ onto $C_{b}(\qr,
\RR)$ is precisely the \emph{Gelfand transformation}. The map $\hr \to
\orr, \ \ A \tto f_{A}$ is therefore for a general von Neumann algebra 
a noncommutative generalization of the Gelfand transformation. 

 \section{The Gelfand Transformation}
\label{sec: BQObc}
\pagestyle{myheadings}
\markboth{Quantum Observables}{The Gelfand Transformation}

Let $\kaa$ be an abelian von Neumann algebra.
Subsequently we will prove that the mapping $A \tto f_{A}$ from $\kaa$
onto $C(\qa, \RR)$ is up to the isomorphism $C(\qa, \RR) \to
C(\gO(\kaa), \RR)$ the Gelfand transformation of the abelian von Neumann
algebra $\kaa$. \\
~\\
Let $\sum_{j = 1}^{m}b_{j}P_{j}$ be an orthogonal representation of $A
\in \lpa$. By lemma 3.6 in \cite{deg3},
\[
    \kf_{\kaa}(A) := \sum_{j = 1}^{m}b_{j}\chi_{\qpja} 
\]
is a well defined continuous function on $\qa$. This defines a mapping
\[
    \kf_{\kaa} : \lpa \to C(\qa).
\]

\begin{proposition}\label{gt1}
    $\kf_{\kaa} : \lpa \to C(\qa)$ is an isometric homomorphism
    of algebras.
\end{proposition}
\emph{Proof:} Let $A = \sum_{i = 1}^{m}a_{i}P_{i}$ and $B = \sum_{j = 
1}^{n}b_{j}Q_{j}$ be orthogonal representations of $A, B \in \lpa$.
Because $\pa$ is distributive we can write
\begin{eqnarray*}
    A + B & = & \sum_{i = 1}^{m}a_{i}P_{i} + \sum_{j = 1}^{n}b_{j}Q_{j}  \\
     & = & \sum_{i}a_{i}(P_{i}Q_{1} + \ldots + P_{i}Q_{n} + P_{i}(I - 
     (Q_{1} + \ldots + Q_{n})))  \\
     &  & + \sum_{j}b_{j}(Q_{j}P_{1} + \ldots + Q_{j}P_{m} + Q_{j}(I -
   (P_{1} + \ldots + P_{m})))  \\
     & = & \sum_{i, j}(a_{i} + b_{j})P_{i}Q_{j}  \\
     &  & + \sum_{i}a_{i}P_{i}(I - (Q_{1} + \ldots + Q_{n})) +
     \sum_{j}b_{j}Q_{j}(I - (P_{1} + \ldots + P_{m})).
\end{eqnarray*}
This is an orthogonal representation for $A + B$. Applying
$\kf_{\kaa}$ gives
\begin{eqnarray*}
    \kf_{\kaa}(A + B) & = & \sum_{i, j}(a_{i} +
    b_{j})\chi_{\kQ_{P_{i}Q_{j}}(\kaa)}  \\
     &  & + \sum_{i}a_{i}\chi_{\kQ_{P_{i}(I - (Q_{1} + \ldots +
     Q_{n}))}(\kaa)} + \sum_{j}b_{j}\chi_{\kQ_{Q_{j}(I - (P_{1} + \ldots +
     P_{m}))}(\kaa)}  \\
     & = & \sum_{i}a_{i}\chi_{\kQ_{P_{i}}(\kaa)\cap \kQ_{Q_{1} + \ldots +
     Q_{n}}(\kaa)} + \sum_{j}b_{j}\chi_{\kQ_{Q_{j}}(\kaa) \cap \kQ_{P_{1} + \ldots 
     + P_{m}}(\kaa)}  \\
     &  & + \sum_{i}a_{i}\chi_{\kQ_{P_{i}}(\kaa) \cap \kQ_{I - (Q_{1} + \ldots +
     Q_{n})}(\kaa)} + \sum_{j}b_{j}\chi_{\kQ_{Q_{j}}(\kaa) \cap \kQ_{I - (P_{1} + \ldots +
     P_{m})}(\kaa)}  \\
     & = & \kf_{\kaa}(A) + \kf_{\kaa}(B).
\end{eqnarray*}
Trivially $\kf_{\kaa}(cA) = c \kf_{\kaa}(A)$ for $A \in \lpa$ and $c
\in \CC$. A simple calculation shows that $\kf_{\kaa}$ is also
multiplicative:
\[
    \kf_{\kaa}(AB) = \kf_{\kaa}(A) \kf_{\kaa}(B).
\]
Let $A = \sum_{i = 1}^{m}a_{i}P_{i}$ be an orthogonal representation
of $A \in \lpa$. Then
\[
    |A| = \max_{i ≤ m}|a_{i}|
\]
and
\[
    |\sum_{i}a_{i}\chi_{\kQ_{P_{i}}(\kaa)}|_{\∞} = \max_{i ≤ m}|a_{i}|
\]
because the sets $\kQ_{P_{i}}(\kaa)$ are pairwise disjoint. Hence
$\kf_{\kaa}$ is isometric. \ \ $\Box$ \\

\begin{corollary}\label{gt2}
    $\kf_{\kaa}$ has a unique extension to an isometric $*$-isomorphism
    from $\kaa$ onto $C(\qa)$. We denote this extension again by $\kf_{\kaa}$.
\end{corollary}
\emph{Proof:} It follows from the Stone-Weierstrass-theorem that
$lin_{\CC}\{ \chi_{\qpa} | P \in \pa \}$ is dense in $C(\qa)$. The
unique isometric extension of $\kf_{\kaa}$ to $\kaa$ is therefore also
surjective. \ \ $\Box$ \\

\begin{proposition}\label{gt3}
    $\kf_{\kaa} : \kaa \to C(\qa)$ is the Gelfand transformation of
    the abelian von Neumann algebra $\kaa$.
\end{proposition}
\emph{Proof:} Let
\[
    \eps_{\gb} : C(\qa) \to \CC
\]
denote the evaluation at the quasipoint $\gb \in \qa$:
\[
    \all \ \gf \in C(\qa) : \ \eps_{\gb}(\gf) = \gf(\gb).
\]
Then for all $P \in \pa$
\begin{eqnarray*}
    (\eps_{\gb} \circ \kf_{\kaa})(P) & = & \eps_{\gb}(\chi_{\qpa})  \\
     & = & \begin{cases}
     1  &  \text{if} \quad P \in \gb \\
     0  &  \text{otherwise}
     \end{cases}
       \\
     & = & \gt_{\gb}(P),
\end{eqnarray*}
hence $\eps_{\gb} \circ \kf_{\kaa} = \gt_{\gb}$ on a dense part of
$\kaa$ and therefore, by continuity, on all of $\kaa$. \\
The Gelfand transformation
\[
    \gG : \kaa \to C(\gO(\kaa)), \quad A \tto \Hat{A},
\]
is defined by
\[
    \all \ \gt \in \gO(\kaa) : \ \Hat{A}(\gt) := \gt(A).
\]
The homeomorphism $\gtt : \gb \tto \gt_{\gb}$ from $\qa$ onto
$\gO(\kaa)$ induces a $*$-isomorphism
\[
    \gtt^{*} : C(\gO(\kaa)) \to C(\qa), \quad \gf \tto \gf \circ \gtt.
\]
We obtain
\[
    \kf_{\kaa} = \gtt^{*} \circ \gG,
\]
because 
\[
    \gtt^{*}(\Hat{A})(\gb) = \Hat{A}(\gtt(\gb)) = \Hat{A}(\gt_{\gb}) =
    \gt_{\gb}(A) = \eps_{\gb}(\kf_{\kaa}(A)) = \kf_{\kaa}(A)(\gb)
\]
holds for all $A \in \kaa$ and all $\gb \in \qa$. In this sense
$\kf_{\kaa}$ ``is'' the Gelfand transformation of $\kaa$. \ \
$\Box$ \\

\begin{theorem}\label{gt4}
    Let $\kaa$ be an abelian von Neumann algebra. Then the mapping
    $A \tto f_{A}$ from $\kaa$ onto $C(\qa, \RR)$ is the restriction 
    of the Gelfand transformation to $\kaa_{sa}$.
\end{theorem}
\emph{Proof:} Due to the foregoing proposition we only need to show
that $f_{A} = \kf_{\kaa}(A)$ holds for all $A \in \kaa_{sa}$. By
corollary \ref{cor: 8} this is true for all $A \in lin_{\RR}\pa$. Let 
$A$ be an arbitrary element of $\kaa_{sa}$. We have seen in the
proof of theorem \ref{theo: 9} that $f_{A}$ is the uniform limit of
observable functions $f_{B}$ with $B \in lin_{\RR}\pa$ and by
definition $\kf_{\kaa}(A)$ is the uniform limit of functions
$\kf_{\kaa}(B)$ with $B \in lin_{\RR}\pa$. Hence $f_{A} = \kf_{\kaa}(A)$.
\ \ $\Box$ \\
~\\
If $A$ is an arbitrary element of the abelian von Neumann algebra
$\kaa$ and $A = A_{1} + iA_{2}$ is its decomposition into selfadjoint 
parts, then the Gelfand transform of $A$ is
\[
    \kf_{\kaa}(A) = \kf_{\kaa}(A_{1}) + i\kf_{\kaa}(A_{2}).
\]
It is therefore natural to define the \emph{complex} observable function
of $A$ as
\[
    f_{A} := f_{A_{1}} + if_{A_{2}}.
\]
This definition can be extended to the elements of an arbitrary von Neumann
algebra $\rr$. \\
~\\
~\\
As an application of our considerations we will characterize
\emph{compact normal} operators by its observable functions. We
assume that the Hilbert space $\kh$ has infinite dimension, for
otherwise there is nothing to characterize. Let $A \in \lh_{sa}$ be compact.
It is well known that $A$ can be represented as
\begin{equation}
    A = \sum_{k \inn}\gl_{k}P_{\CC e_{k}},
    \label{eq:9}
\end{equation}
where $\{ e_{k} \ | \ k \inn \}$ is a maximal orthonormal set of
eigenvectors and the sequence $(\gl_{k})_{k \inn}$ of eigenvalues
converges to zero. The sum converges with respect to the norm. \\
Now let $\mm$ be a maximal abelian von Neumann subalgebra of $\lh$
corresponding to a maximal atomic Boolean sector of $\ph$ such that
$\mm$ contains $A$ and the projections $P_{k} := P_{\CC e_{k}}$ 
for all $k \inn$. Consider the finite-rank approximation
\[
    A_{n} := \sum_{k = 1}^{ n}\gl_{k}P_{k}
\]
of $A$. The observable function of $A_{n}$ is
\[
    f_{A_{n}} = \sum_{k = 1}^{n}\gl_{k}\chi_{\qpka}  
     = \sum_{k = 1}^{n}\gl_{k}\chi_{\{\gb_{P_{k}}\}},
\] 
where $\gb_{P_{k}} \in \qmm$ denotes the atomic quasipoint defined by
$P_{k}$. This means that $f_{A_{n}}$ has finite support, contained in 
$\{ \gb_{P_{1}}, \ldots, \gb_{P_{n}} \}$. In particular, $f_{A_{n}}$
vanishes on the closed set $\qmm_{c}$ of \emph{continuous} (i.e.
non-atomic) quasipoints of $\pmm$. Since the functions $f_{A_{n}}$ are 
the Gelfand transforms of the operators $A_{n}$ and since the sequence
$(A_{n})_{n \inn}$ converges in norm to $A$, the sequence
$(f_{A_{n}})_{n \inn}$ converges \emph{uniformly} to the observable
function $f_{A}$ of $A$. Hence $f_{A}$ vanishes on $\qmm_{c}$ and,
considered as a function on the open set $\qmm_{at}$ of atomic quasipoints
of $\pmm$, is an element of $C_{0}(\qmm_{at})$, the algebra of continuous
functions $\qmm_{at} \to \CC$ that vanish at infinity. Note that
$\qmm_{at}$ is an open discrete and dense subspace of $\qmm$.
Therefore $\qmm_{c}$ is the boundary of $\qmm_{at}$.
\\
~\\
Conversely, let $f : \qmm \to \RR$ be a continuous function that satisfies
\begin{enumerate}
    \item  [(i)] $f_{|_{\qmm_{c}}} = 0$ and

    \item  [(ii)] $f_{|_{\qmm_{at}}} \in C_{0}(\qmm_{at})$
\end{enumerate}
Then $f$ is the uniform limit of a sequence $(f_{n})_{n \inn}$ of
functions $f_{n} : \qmm \to \RR$ of finite support contained in
$\qmm_{at}$. The selfadjoint operator $A_{n} \in \mm$ is therefore a
finite real linear combination of rank-one projections and hence of
finite rank. The sequence $(A_{n})_{n \inn}$ converges in norm to the 
selfadjoint $A \in \mm$ that corresponds to $f$. Hence $f$ is the
observable function of the \emph{compact} selfadjoint operator $A$. \\

\noindent{Conditions} $(i), (ii)$ are not independent: we show that
$(i)$ implies $(ii)$ in a quite general situation.

\begin{lemma}\label{gt5a}
     Let $M$ be a compact Hausdorff space, $D \tm M$ the (discrete
     open) set of isolated points of $M$ and $X := M \smm D$. If $f
     \in C(M)$ vanishes on $X$, then $f$ vanishes at infinity on $D$.  
\end{lemma}
\emph{Proof:} We assume that $D$ is an infinite set, for otherwise
there would be nothing to prove. Let $f : M \to \CC$ be a continuous
function that vanishes on $X$. If $\eps > 0$, we can choose for every 
$x \in X$ an open neighbourhood $U_{x}$ of $x$ such that $|f(y)| ≤
\eps$ for all $y \in U_{x}$. The open sets $U_{x} \ (x \in X)$
together with the open sets $\{p\} \ (p \in D)$ form an open covering 
of the compact space $M$. Hence there are only finitely $p_{1},
\ldots, p_{n} \in D$ that do not belong to $\bigcup_{x \in X}U_{x}$.
This means $|f(p)| ≤ \eps$ for all $p \in D \smm \{p_{1}, \ldots,
p_{n}\}$, i.e. $f$ vanishes at infinity on $D$. \ \ $\Box$ \\

\noindent{If} the set $D$ of isolated points of $M$ is dense in $M$,
then we can show that condition $(ii)$ implies condition $(i)$:

\begin{lemma}\label{gt5b}
    Let $M$ be a compact Hausdorff space, $D \tm M$ the (discrete
    open) set of isolated points of $M$ and $X := M \smm D$. If $D$ is
    dense in $M$, then every $f \in C(M)$ that vanishes at infinity on
    $D$, vanishes on the boundary $X$ of $D$.
\end{lemma}
\emph{Proof:} Again we can assume that $D$ is an infinite set. Let 
$x \in X$ such that $f(x) \ne 0$. We may assume that $f(x) = 1$. Let
$U_{1}$ be an open neighbourhood of $x$ such that $|f(y)| ≥
\frac{1}{2}$ for all $y \in U_{1}$. Since $\overline{D} = M$, there is
some $p_{1} \in U_{1} \cap D$. Choose a neighbourhood $U_{2}$ of $x$
that is contained in $U_{1}$ and does not contain $p_{1}$. Then choose
$p_{2} \in U_{2} \cap D$. Proceeding in this way, we generate a
sequence $(p_{n})_{n \inn}$ of infinitely many different points in $D$
such that $|f(p_{n})| ≥ \frac{1}{2}$ for all $n \inn$. Hence $f$ does 
not vanish at infinity on $D$. \ \ $\Box$ \\ 
~\\
The set $\kK(\omm)$ of all continuous functions $f : \qmm \to \CC$
that vanish on $\qmm_{c}$ forms a \emph{selfadjoint} ideal in
$C(\qmm)$. Also the set $\kK(\mm)$ of all compact operators in
$\mm$ is a selfadjoint ideal.   
Summing up, we have proved the following

\begin{proposition}\label{gt6}
    Let $\kh$ be an infinite dimensional Hilbert space and let $\mm$ be a
    maximal abelian von Neumann subalgebra of $\lh$ corresponding to a
    maximal atomic Boolean sector of $\ph$. Let $\qmm_{at}$ be the open
    discrete (and dense) set of atomic quasipoints of $\pmm$ and let
    $\qmm_{c} := \qmm \smm \qmm_{at}$ be the set of continuous
    quasipoints. Then the restriction of the Gelfand transformation
    $\kf_{\mm} : \mm \to \omm$ to the ideal $\kK(\mm)$ of all compact
    operators in $\mm$ is an isometric isomorphism from $\kK(\mm)$ onto
    the ideal $\kK(\omm)$ in $\omm$ of all $f \in C(\qmm)$ that vanish
    on $\qmm_{c}$ (or, equivalently, vanish at infinity on
    $\qmm_{at}$).
\end{proposition}

\end{document}